\documentclass[5p]{elsarticle}
\biboptions{sort&compress} 
\usepackage{graphics}
\usepackage{amssymb}
\usepackage{lineno}
\usepackage{array}
\usepackage{subfigure}
\usepackage{ifthen}
\usepackage{wasysym}
\usepackage{multirow}
\usepackage[hyperindex,breaklinks]{hyperref}

\usepackage{preprintcover}  
\PreprintCoverPaperTitle{Search for the decay \Bs$ \to\! \mu^+\! \mu^-$ with the ATLAS detector}  
\PreprintIdNumber{CERN-PH-EP-2012-067}  
\PreprintCoverAbstract{A blind analysis searching for the decay \Bs$ \to \mumu $ has been performed using  proton-proton collisions at a centre-of-mass energy of 7 TeV recorded with the ATLAS detector at the LHC.  With an integrated luminosity of 2.4 fb$^{-1}$ no excess of events over the background expectation is found and an upper limit is set on the branching fraction   BR(\Bs $\to \mumu) < 2.2 (1.9) \times 10^{-8}$ at  95\% (90\%) confidence level.
}  
\PreprintJournalName{Physics Letters B}  

\let\theversion=\relax
\newcommand{\theversion}{0.0}
\newcommand{\version}[1]{\renewcommand{\theversion}{#1}}

\def\TeV{\ifmmode {\mathrm{\ Te\kern -0.1em V}}\else
                   \textrm{Te\kern -0.1em V}\fi}%
\def\GeV{\ifmmode {\mathrm{\ Ge\kern -0.1em V}}\else
                   \textrm{Ge\kern -0.1em V}\fi}%
\def\MeV{\ifmmode {\mathrm{\ Me\kern -0.1em V}}\else
                   \textrm{Me\kern -0.1em V}\fi}%
\def\keV{\ifmmode {\mathrm{\ ke\kern -0.1em V}}\else
                   \textrm{ke\kern -0.1em V}\fi}%
\def\eV{\ifmmode  {\mathrm{\ e\kern -0.1em V}}\else
                   \textrm{e\kern -0.1em V}\fi}%

\newcommand{\Bs}{\ensuremath{B^0_s}}

\newcommand{\pt}{\ensuremath{p_{\mathrm T}}}
\newcommand{\Jmumu}{\ensuremath{J/\psi \to \mumu}}

\newcommand{\mumu}{\ensuremath{\mu^+\mu^-}}
\newcommand{\earatio}{\ensuremath{{R}_{A\epsilon}}}
\newcommand{\thetau}{\ensuremath{{R}^{\rm bkg}}}
\newcommand{\Bsmumu}{\ensuremath{B_s^0 \to \mu^+\mu^-}}

\newcommand{\BpmKpmJpsi}{\ensuremath{B^\pm \to J/\psi K^\pm}}
\newcommand{\BsJpsiPhi}{\ensuremath{B_s^0 \to J/\psi\phi}}

\newcommand{\theRho}{\ensuremath{\frac{A_{J/\psi K^\pm}}{A_{\mu^+\mu^-}}\frac{\epsilon_{J/\psi K^\pm}}{\epsilon_{\mu^+\mu^-}}}}
\newcommand{\Pythia}{\textsc{Pythia}}

\newcommand{\Geant}{\textsc{Geant4}}

\newboolean{extramaterial}
\setboolean{extramaterial}{false}
\newcommand{\addmaterial}[1]{\protect 
  \ifthenelse{\boolean{extramaterial}}{#1}{}}

\newcommand{\csccomment}[1]{\protect 
  \ifthenelse{\boolean{extramaterial}}{{\em#1}}{}}
\newboolean{convenorcomments}
\setboolean{convenorcomments}{false}
\newcommand{\concomment}[1]{\protect 
  \ifthenelse{\boolean{convenorcomments}}{{\em To convenors: #1}}{}}
\newboolean{reviewercomments}
\setboolean{reviewercomments}{true}
\newcommand{\revcomment}[1]{\protect 
  \ifthenelse{\boolean{reviewercomments}}{{\em To reviewers: #1}}{}}

\newcommand{\B} {\ensuremath{B}}

\version{4.17}
\journal{Physics Letters B}
                                                                                                                                                                                                     
\begin{document}

\begin{frontmatter}

\title{                                                                                                                                                                                                                             
Search for the decay \Bs$ \to\! \mu^+\! \mu^-$ with the ATLAS detector }

\begin{abstract}
A blind analysis searching for the decay \Bs$ \to \mumu $ has been performed using  proton-proton collisions at a centre-of-mass energy of 7 TeV recorded with the ATLAS detector at the LHC.  With an integrated luminosity of 2.4 fb$^{-1}$ no excess of events over the background expectation is found and an upper limit is set on the branching fraction   BR(\Bs $\to \mumu) < 2.2 (1.9) \times 10^{-8}$ at  95\% (90\%) confidence level. 
\end{abstract}

\end{frontmatter}

\section{Introduction} 
Flavour changing neutral current processes are highly suppressed in the Standard Model (SM), 
and therefore their study is of particular interest in the search for new physics. 
The SM predicts the branching fraction for the decay $\Bs\to \mumu$  to be extremely small: $(3.5\pm0.3) \times 10^{-9}$ \cite{Buras:2010zm,Buras:2010wr,utfitpaper,Charles:2004jd}.
This process might be substantially enhanced by coupling to non-SM heavy particles, such as those predicted by 
the Minimal Supersymmetric Standard Model 
\cite{Hall:1993gn,Hamzaoui:1998nu,Babu:1999hn,Choudhury:2005rz,Parry:2005fp,Ellis:2006jy,Ellis:2007kb} and other extensions \cite{Davidson:2010uu}.  
Upper limits on this branching fraction, in the range (0.45--5.1)$\times 10^{-8}$, have been reported by the D0 \cite{Abazov:2010fs}, CDF \cite{Aaltonen:2011fi}, 
CMS \cite{Chatrchyan:2011kr,CMS_2012} and LHCb \cite{LHCb:2011ac,LHCb_2012} collaborations. This Letter reports the result of a search 
performed with $pp$ collisions corresponding to an integrated luminosity of 2.4 fb$^{-1}$, 
collected in the first half of the 2011 data-taking period using the ATLAS
detector at the LHC.

\label{sec:outline} 
The analysis is based on events selected with a di-muon trigger and reconstructed in the ATLAS inner tracking detector and muon spectrometer~\cite{AtlasDet}. Details of the detector, trigger and datasets are discussed in Section~\ref{sec:det-prel}, together with the preselection criteria.

The \Bsmumu\ branching fraction is measured with respect to a prominent reference decay ($B^\pm \to J/\psi K^\pm$) in order to minimize systematic uncertainties in the evaluation of the 
efficiencies and acceptances, while still providing small statistical uncertainties.
The branching fraction can be written as
\begin{eqnarray}
\hbox{\hspace{-1cm}} \mathrm{BR} ( \Bs \to \mumu)=&\hbox{\hspace{-3mm}}\mathrm{BR} (B^\pm \to J/\psi K^\pm\! \to \mumu K^\pm)\times \nonumber \\
                                                       &\frac{f_{u}}{f_{s}} 
\times \frac{N_{\mumu}}{N_{J/\psi K^\pm}}
\times\theRho\, ,\label{BRFormula}
\end{eqnarray}
where the right-hand side includes the $B^\pm \to J/\psi K^\pm\! \to \mumu K^\pm$ branching fraction, the relative 
production probability of $B^\pm$ and \Bs\
$f_u/f_s$ taken from previous measurements \cite{pdg_book, Asner:2010qj, Aaij:2011jp}, the event yields after background subtraction, and the acceptance and efficiency ratios. The event yields for both signal and reference channels were obtained from signal and sideband (background) regions defined in the invariant mass spectrum (see Table \ref{table:MassRegions}).

The Single Event Sensitivity (SES) corresponds to the $\Bs \to\mumu$ branching fraction which would yield one observed signal event
in the data sample:
\begin{equation}
{\rm BR}(\Bs \to \mumu) = N_{\mumu} \times \mathrm {SES}\, ,  \label{SES}
\end{equation} 
where $N_{\mumu}$ is the number of observed events. 

This Letter describes the results of a blind analysis in which the di-muon mass region 5066 to 5666 MeV was removed from the analysis until the procedures for event selection, signal and limit extractions were fully defined.Ê Sections~\ref{subsec:background} to \ref{subsec:MCtuning} discuss  the variables used in the event selection, Monte Carlo (MC) tuning and background studies.
The final sample of candidates was selected with a multivariate classifier, trained on a fraction of the events from the di-muon invariant mass sidebands, 
as discussed in Section~\ref{subsec:TMVA-1}.
The relative efficiency and event yields in the reference channel are discussed in Sections~\ref{subsec:rho} and \ref{subsec:Bplus}, respectively.
The signal extraction is discussed in Section~\ref{sec:extraction} and the corresponding limit on the branching fraction 
 is presented in Section~\ref{sec:limits}.

According to the SM, the branching fraction BR($\B^0 \to \mumu$) is predicted to be about  30 times smaller than BR($\Bs \to \mumu$) \cite{Buras:2010zm,Buras:2010wr}. 
Therefore, despite the increased SES of approximately a factor four due to the absence of the factor $f_u/f_s$ and possible enhancements due to new physics, the sensitivity to this channel is
beyond the reach of the current analysis. Hence only a limit on BR($\Bs \to \mumu$) was derived by assuming BR($\B^0 \to \mumu$) to be negligible.

\section{ATLAS detector, data and simulation samples}\label{sec:det-prel}
The ATLAS detector\footnote{ATLAS uses a right-handed coordinate system with its origin at the nominal interaction point. The $z$-axis is along the beam pipe, the $x$-axis points to the centre of the LHC ring and the $y$-axis points upward. Cylindrical coordinates $(r,\phi)$ are used in the transverse plane, $\phi$ being the azimuthal angle around the beam pipe. The pseudorapidity $\eta$ is defined as $\eta = - {\rm ln}[{\rm tan}(\theta/2)]$ where $\theta$ is the polar angle.}
 consists of three main components: an  Inner Detector tracking system (ID) immersed in
a 2~T magnetic field, a system of electromagnetic and hadronic calorimeters, and an outer 
Muon Spectrometer (MS). A full description can be found in~\cite{AtlasDet}. 
The detector performance characteristics most relevant to this analysis are the vertex-finding and the overall track reconstruction in the ID and MS, together with the ability of the trigger system to record events containing pairs of muons.

The ID provides precise track reconstruction within the pseudorapidity range
 $|\eta| < 2.5$. It employs a Pixel detector close to the beam-pipe, a
silicon microstrip detector (SCT) at intermediate radii and a Transition Radiation Tracker~(TRT) at
outer radii.
The innermost Pixel layer 
is located at a radius of 50.5~mm and plays a key role in precise vertex determination.

The MS comprises separate trigger and high-precision tracking chambers that measure the deflection of muons in a toroidal magnetic field. 
The precision chambers cover the region $|\eta| < 2.7$ and measure the coordinate in the bending plane.
The trigger chambers cover the range $|\eta| < 2.4$ and provide fast coarser measurements in both the bending 
and non-bending plane. 

This analysis is based on a sample of $pp$ collisions at $\sqrt{s}=7$~TeV, recorded by ATLAS in the period April--August  2011. 
Trigger and pile-up conditions changed for data taken after this period: the remainder of the 2011 dataset will be included in a future analysis.
Data used in the analysis were recorded during stable LHC beam periods. Further data quality requirements were also imposed, notably on the performance of the MS and ID systems.
The total integrated luminosity amounts to  2.4~fb$^{-1}$. This sample has an average of about five primary vertices per event
from multiple proton-proton interactions.

A muon trigger \cite{Aad:2012xs} was used to select events.
In particular, the sample contains events seeded by a Level-1 di-muon trigger 
which required a transverse momentum \pt$>$4~\GeV\ for both muon candidates. A full track reconstruction of the muon candidates was performed at the second and third trigger levels, where additional cuts on the di-muon invariant mass $m_{\mumu}$ were applied, loosely selecting events compatible with $J/\psi$ (2500 to 4300 \MeV) or  \Bs\  (4000 to 8500 \MeV) decays into a muon pair. 

Events containing candidates for \Bs$\to \mumu$, $B^\pm \to J/\psi K^\pm \to \mumu K^\pm$ and, as discussed in Sections~\ref{subsec:discr-var} and \ref{subsec:MCtuning},
\Bs$ \to J/\psi \phi \to \mumu K^+ K^-$ were retained for this analysis. After cutting on the mass of the intermediate resonances (1009 \MeV $\le m_\phi \le$ 1031 \MeV, 2915 \MeV $\le m_{J/\psi} \le$ 3175 \MeV)  a preselection was applied, based on track properties and the 
quality of the reconstructed $B$ decay vertex.  All charged particle tracks reconstructed in the ID were required to have at least one Pixel, six SCT and eight TRT hits.
Tracks were required to have $|\eta|<2.5$ and $\pt>4$~\GeV\ ($>2.5$~\GeV) for muon (kaon) candidates.
No particle identification was used to distinguish $K^\pm$ and $\pi^\pm$ candidates.
ID tracks that were matched to reconstructed MS tracks were selected as candidate muons.
Decay vertices were formed by combining two, three or four tracks, according to the specific decay process \cite{mc10detail}.  
All $B$ meson properties were computed based on the result of the fit of the tracks to the $B$ decay vertex. In order to reject fake track combinations,
 the fit $\chi^2$ per degree of freedom was required to be less than 2.0 (85\% efficient) for  \Bs$\to \mumu$ and less than 6.0 (99.5\% efficient) for the other channels.
All reconstructed $B$ candidates were required to satisfy $\pt^B>8.0$~\GeV\ and $|\eta^B|<2.5$ in order to
define our efficiencies and acceptances within a fiducial phase-space volume with as little as possible reliance on MC extrapolations.
Signal and sideband regions were defined according to Table \ref{table:MassRegions}.

\begin{table}[htdp]
\begin{center}
\begin{tabular}{l|c|c}
\hline
Channel & Signal Region & Sideband Regions \\
\hline\hline
\multirow{2}{*}{$\Bs\to\mumu$}                 & \multirow{2}{*}{[5066,5666] \MeV} & [4766,5066] \MeV\\ &&[5666,5966] \MeV\\
\hline
\multirow{2}{*}{$B^\pm\to J/\psi K^\pm$} & \multirow{2}{*}{[5180,5380] \MeV} & [4930,5130] \MeV\\ &&[5430,5630] \MeV \\
\hline\hline
\end{tabular}
\end{center}
\caption{Definition of the signal and sideband regions used in this analysis.}
\label{table:MassRegions}
\end{table}%

The primary vertex position was obtained from a fit of charged tracks not used in the decay vertex and constrained to the interaction region of the colliding beams.
If multiple candidate primary vertices were present, the one closest in $z$ to the decay vertex was chosen.
After preselection, approximately 2$\cdot 10^5$ $B^0_s \to \mumu$ and 1.4$\cdot 10^5$ $B^\pm \to J/\psi K^\pm$ candidates were obtained
in the signal regions.

Samples of Monte Carlo (MC) events were used for the extraction of acceptance and efficiency ratios. MC samples were produced for the signal channel $\Bs \to \mumu$, the reference channel \BpmKpmJpsi\  (\Jmumu) and the control channel $\Bs \to J/\psi \phi$ ($\phi \to K^+K^-$). These samples were generated with \Pythia\ 6.4~\cite{Sjostrand:2006za} using the 2010 ATLAS~\cite{mctunes,mc10detail} tune. 
MC events were filtered before detector simulation to ensure the presence of at least one decay of interest, with B decay products all satisfying  $|\eta|<2.5$ and $\pt >2.5$ (0.5) \GeV\ for muons (kaons). An additional sample was generated with a fictitious value of the \Bs\ mass (6500~\MeV) and the same parameters as the standard \Bsmumu sample, allowing a check of the full analysis on a signal-free region before unblinding.
The ATLAS detector and its response were simulated using \Geant~\cite{GEANT4}.
Additional $pp$ interactions in the same and nearby bunch crossings (pile-up) were included in the simulation. 

\section{Event selection} \label{sec:selection} 
This Section describes the expected background composition, the discriminating variables used as input to the multivariate classifier, the tuning of the simulation for the determination of the signal efficiency, the data samples used to estimate the background rejection and the optimization procedure.
The signal efficiency was determined from MC samples, re-weighted to account for differences between data and MC simulation of the $B$ meson kinematics. The rejection power was tested using a sub-sample of background events from the sidebands in the di-muon mass spectrum.

\subsection{Background composition}\label{subsec:background}
Two categories of background were considered: a continuum with a smooth dependence on the di-muon invariant mass, 
and sources of resonant contributions from mis-reconstructed decays.

Comparisons of data and MC have shown that the combinatorial background  from  $b\bar{b} \to  \mumu X$ decays provides a reasonable description
for the distributions of the discriminating variables for the events found in the sidebands. The $b\bar{b} \to  \mumu X$ MC
sample used is equivalent to about $12$ pb$^{-1}$ of integrated luminosity.
Such studies support the procedure of modeling the continuum background through interpolation of the di-muon yield in the sidebands, but do not reach a sufficient statistical precision. 
Half of the data events in the sidebands (those with odd event numbers) were used to optimize the selection procedure.  The remaining events were used
for the measurement of the background and for interpolation to the signal region.   

Resonant background is due to $B$ decay candidates containing either one or two hadrons erroneously identified as muons. 
Mis-identification may be due to punch-through of a hadron to the MS or to decays in flight where the muon carries most of the hadron momentum.  
In either case the hadron {\it fakes} the muon signature for the purpose of this analysis.
Single-fake events are due to, e.g. $\Bs\to K^+\mu^-\nu$, the charged $K$ meson being mis-identified as a muon. Double-fake
events are due to two-body hadronic $B$ decays ($B \to hh$), e.g. $\Bs \to K^+\pi^-$, where both hadrons are mis-identified as muons. 
MC studies have shown that double-fake events are the main source of resonant background after the selection criteria used in this analysis. 
The main contribution is from $\Bs \to K^+K^-$, followed by $\B^0 \to \pi^+\pi^-$ and $B^0\to K^\pm\pi^\mp$ \cite{pdg_book,Aaltonen:2011jv}. 

The simulation determined the probability for a hadron to be misidentified as a muon to be equal to  $2 \, (4) \, \permil$ for $\pi^\pm$ ($K^\pm$), with a relative uncertainty of $20\%$, validated against control samples in data \cite{ConfNoteFromPeter}.
The value for charged $K$ mesons was averaged over $K^+$ and $K^-$ and was found consistent with the preliminary results of data-driven studies based on the decay  $D^{*}\to D^0\pi\to K\pi\pi$.  

The expected event yield for $B \to hh$ was obtained from an estimation of the integrated luminosity, acceptance and efficiency.
This constitutes a nearly irreducible background in this analysis, due to its resemblance to the actual signal.

\begin{table}[htb!]  
\begin{center}
\footnotesize
\begin{tabular}{ll}
\hline
 Variable & Description 	\vbox{\vspace{3mm}}\\
\hline
\hline \vspace{1mm}
 $|\alpha_{2\mathrm D}|$ & Absolute value of the angle in the transverse\vbox{\vspace{3mm}} \\
 pointing angle &plane between $\Delta \vec x$ and $\vec p{}^B$\\
\hline\vspace{1mm}
 $\Delta R$ & Angle $\sqrt{(\Delta \phi{})^2+(\Delta \eta{})^2}$ between $\Delta\vec x$ and $\vec p{}^B$ \vbox{\vspace{3mm}}\\
\hline\vspace{1mm}
 \multirow{2}{*}{$L_{\rm xy}$} & Scalar product in the transverse plane of\vbox{\vspace{3mm}} \\ & $(\Delta\vec x \cdot \vec p^B)/|\vec p^B_{\rm T}|$ \\ \vspace{-3mm} & \\
\hline\vspace{1mm}
 \multirow {2}{*}{$ct$ significance}
    & Proper decay length $ct=L_{\rm xy}\times m_B / p_{\mathrm T}^B$ \vbox{\vspace{3mm}} \\ \vspace{1mm}
    & divided by its uncertainty\\ 
\hline
\multirow{2}{*}{$\chi^2_{\rm xy}$, $\chi^2_{\rm z}$} & Vertex separation significance $\Delta\vec{x}^{\,T} \! \cdot \left(\sigma_{\Delta\vec{x}}^2\right)^{-1} \! \cdot \Delta\vec{x}$  \vbox{\vspace{3mm}}\\ \vspace{1mm}
   &  in $(x,y)$ and $z$, respectively \\
\hline
      & Ratio of $|\vec p^B_{\rm T}|$ to the sum of $|\vec p^B_{\rm T}|$ and  \vbox{\vspace{3mm}} \\
 $I_{0.7}$      & the transverse momenta of all tracks with  \\ 
 isolation    & $p_{\rm T}>0.5$ \GeV\ within a cone $\Delta R<0.7$ from \\ \vspace{1mm}
      & the $B$ direction, excluding $B$ decay products  \\
\hline
 \multirow {4}{*}{$|d_{\rm 0}^\mathrm {max}|$,  $|d_{\rm 0}^\mathrm {min}|$}
         & Absolute values of the maximum and\vbox{\vspace{3mm}} \\
         &  minimum impact parameter in the\\ 
         & transverse plane of the $B$ decay products\\
         &  relative to the primary vertex\\
\hline
 \multirow {3}{*}{$|D_{\rm xy}^{\rm min}|$, $|D_{\rm z}^{\rm min}|$}
      & Absolute values of the minimum distance of \vbox{\vspace{3mm}} \\ 
      &  closest approach in the $xy$ plane (or along $z$)\\ \vspace{1mm}
      & of tracks in the event to the $B$ vertex\\
\hline\vspace{1mm}
 $p_{\mathrm T}^B$& $B$ transverse momentum \vbox{\vspace{3mm}}  \\
\hline
  \multirow{2}{*} {$p_{\rm L}^\mathrm{max}$, $p_{\rm L}^\mathrm{min}$ }
   &  Maximum and minimum momentum of the two\vbox{\vspace{3mm}}  \\ \vspace{1mm}
    & muon candidates along the $B$ direction\\
\hline
\hline
\end{tabular}
\end{center}
\caption{List of the discriminating variables used in this analysis to separate $B^0_s\to\mumu$ signal from backgrounds.
These variables are based on properties of the decay products, of the reconstructed primary ($\vec{x}_{\rm PV}$) and secondary ($\vec{x}_{\rm SV}$) vertices (separated by $\Delta\vec x=\vec{x}_{\rm SV}-\vec{x}_{\rm PV}$), the $B$ meson momentum $\vec{p}^B$ and the properties of additional tracks from the underlying event. Variables are listed in order of relevance as ranked by the multivariate classifier used in the final signal/background separation as discussed in Section \ref{subsec:TMVA-2}. 
\label{table:discri-var}
}
\end{table}

\subsection{Discriminating variables} \label{subsec:discr-var}

Table~\ref{table:discri-var} describes the discriminating variables used in the multivariate classifier. 
The \Bsmumu\ signal is characterized by the separation between the production (primary) and decay (secondary) vertices, as well
as the two-body decay topology.
These variables exploit such features to discriminate against potential
backgrounds: pairs of prompt charged tracks (e.g. $L_{\rm xy}$, $ct$ significance, $\chi^2_{\rm xy}$), as well as pairs of displaced muons originating from $b\bar{b} \to  \mumu X$ processes (e.g. $d_0^{\rm max}$, $d_0^{\rm min}$),
secondary vertices with additional particles in the final state (e.g. $\alpha_{2D}$, $\Delta R$, $D^{\rm min}_{\rm xy}$, $D^{\rm min}_{z}$) and non-$b\bar{b}$ processes (e.g. $I_{0.7}$, $p_{\rm T}^B$, $p_{\rm L}^{\rm max}$, $p_{\rm L}^{\rm min}$).

\begin{figure*}[htb!] 
\center
\includegraphics[width=0.25\linewidth]{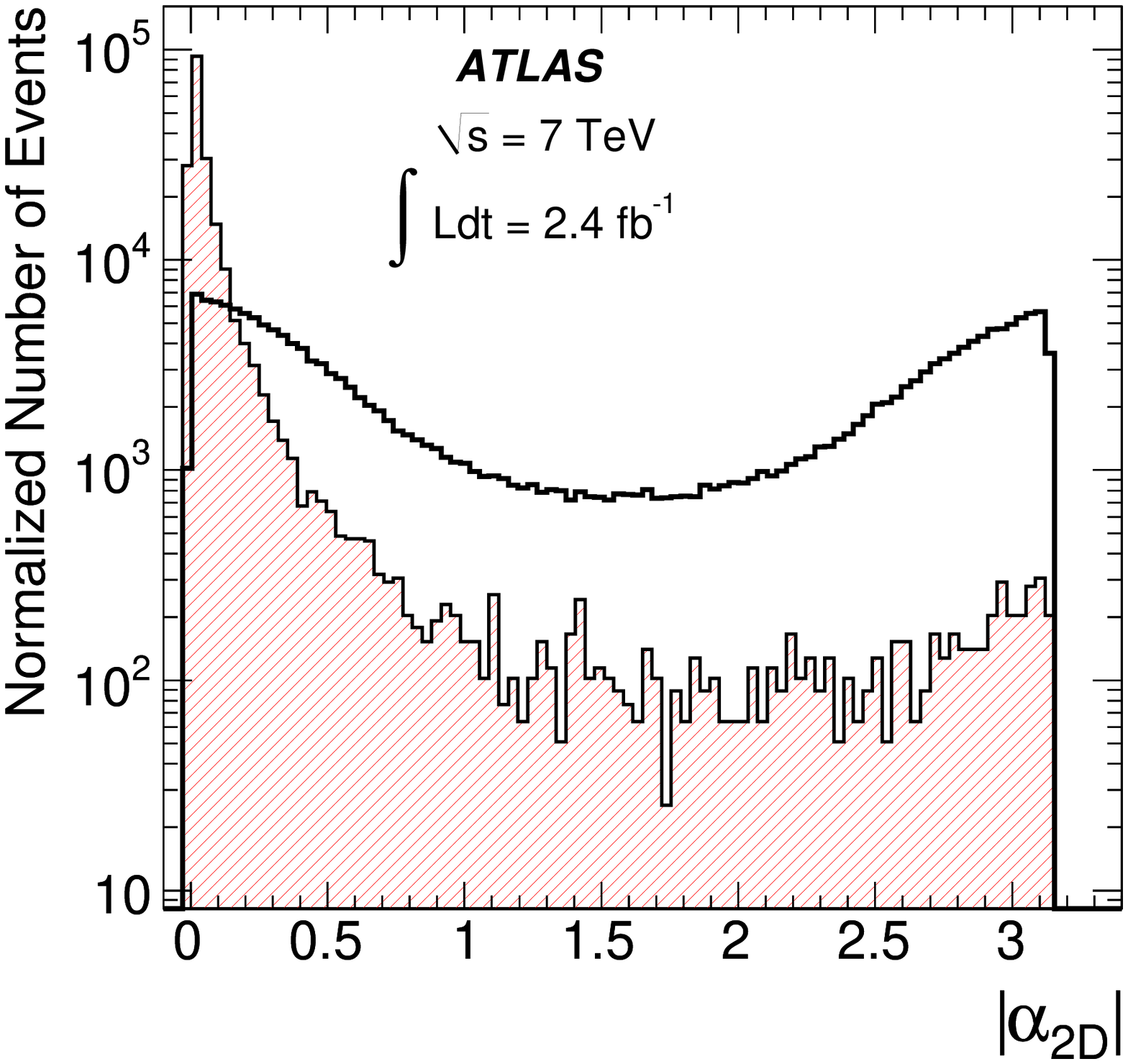}
\includegraphics[width=0.25\linewidth]{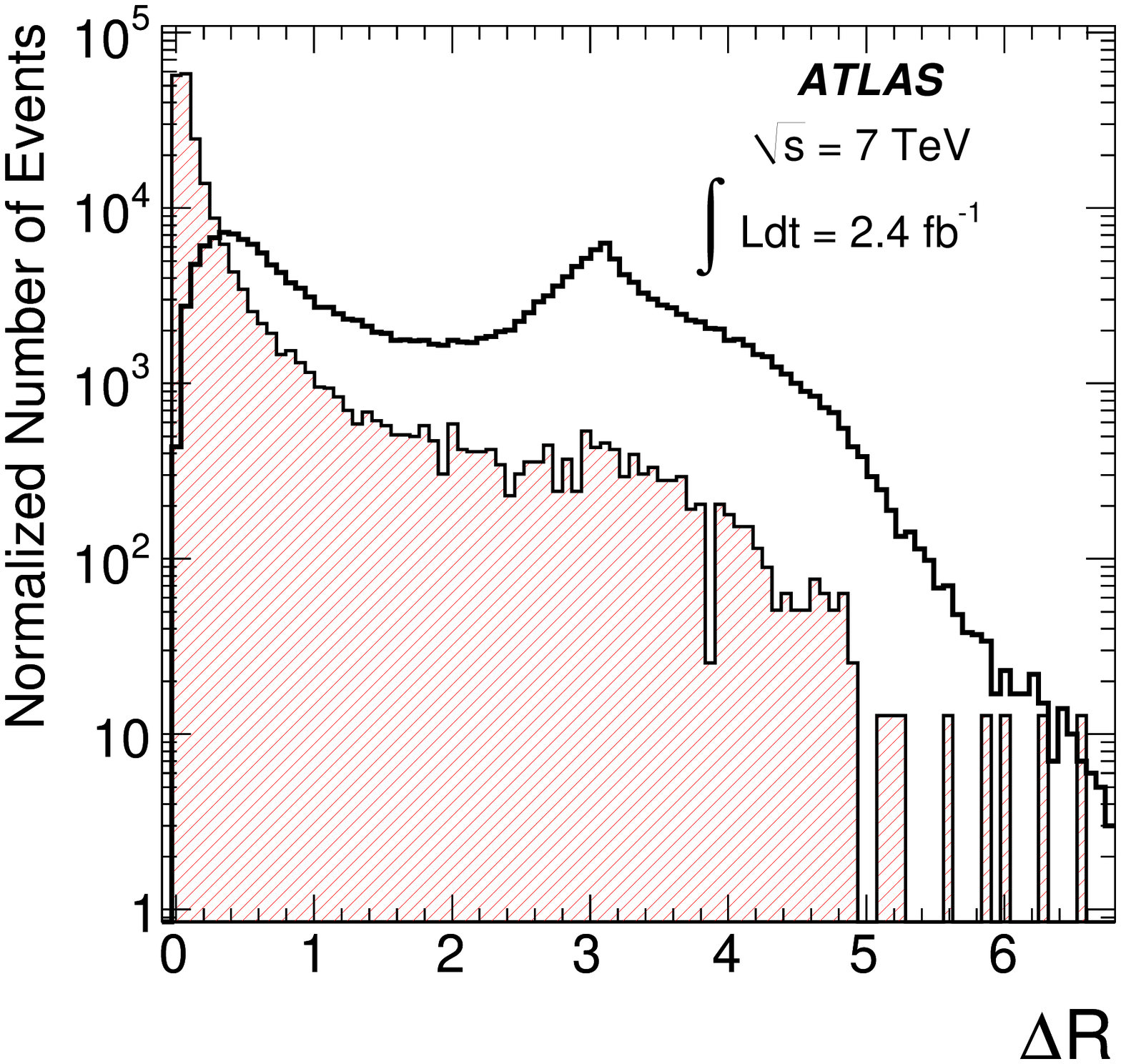} 
\includegraphics[width=0.25\linewidth]{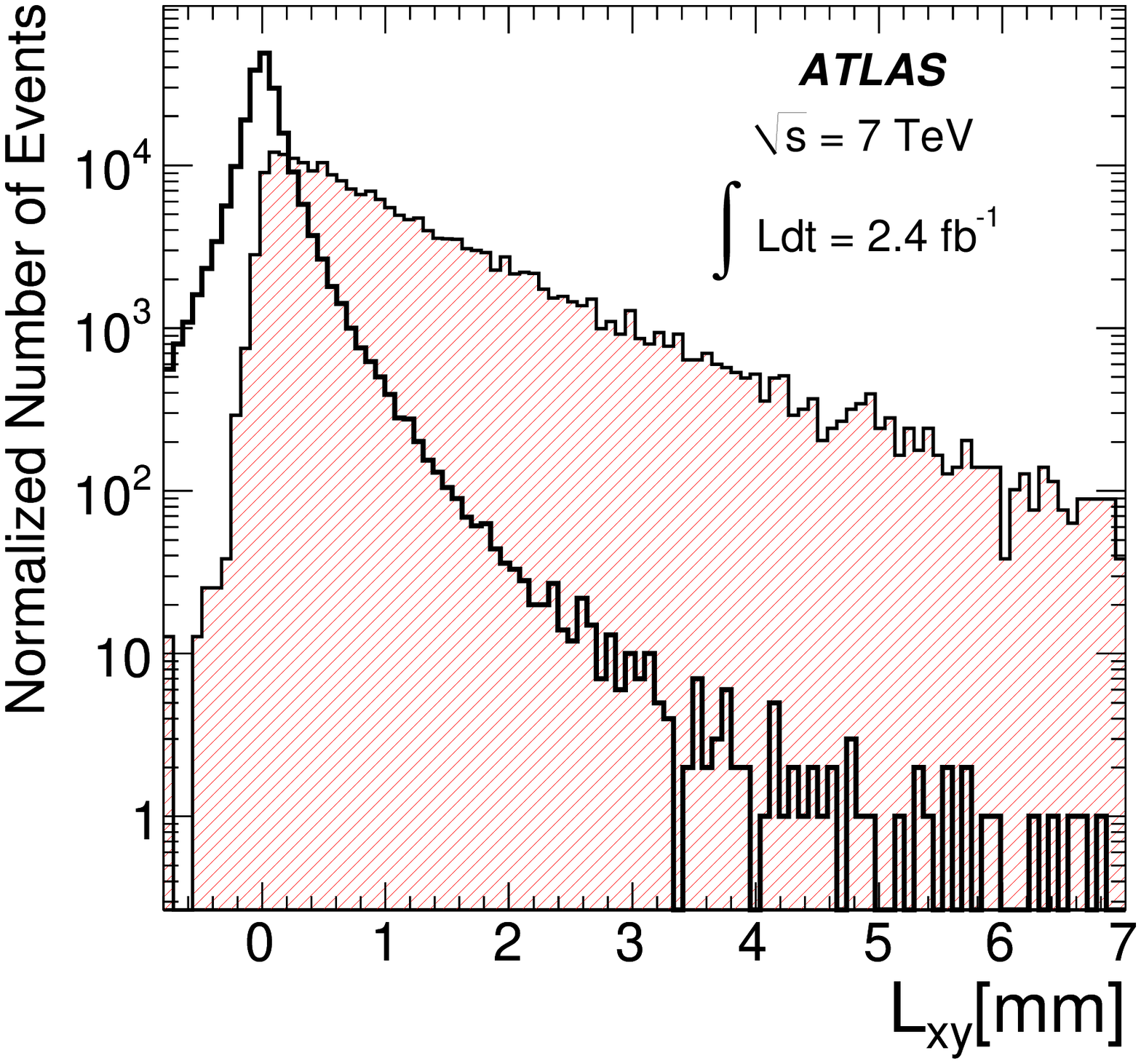} \\ 
\includegraphics[width=0.25\linewidth]{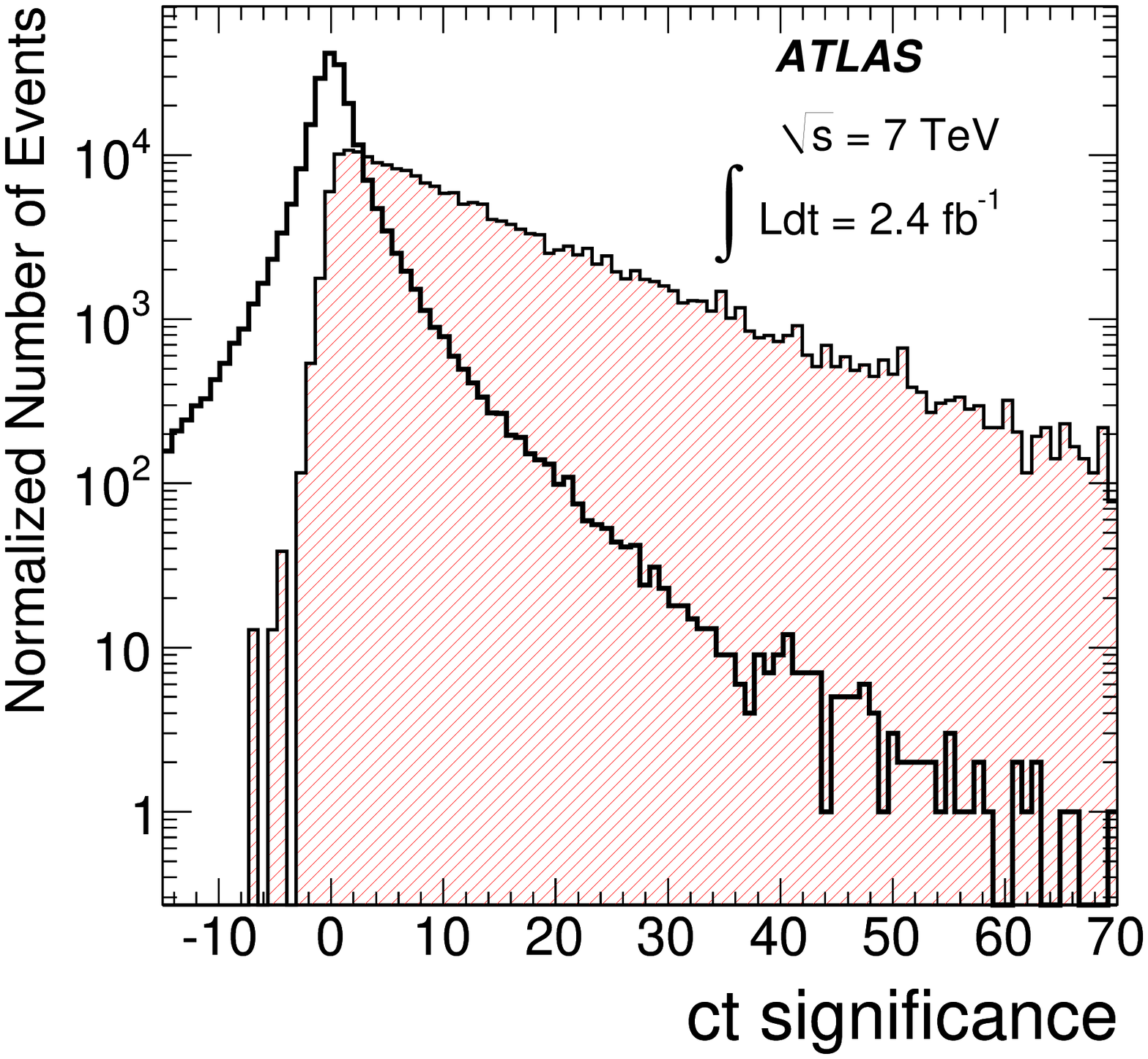} 
\includegraphics[width=0.25\linewidth]{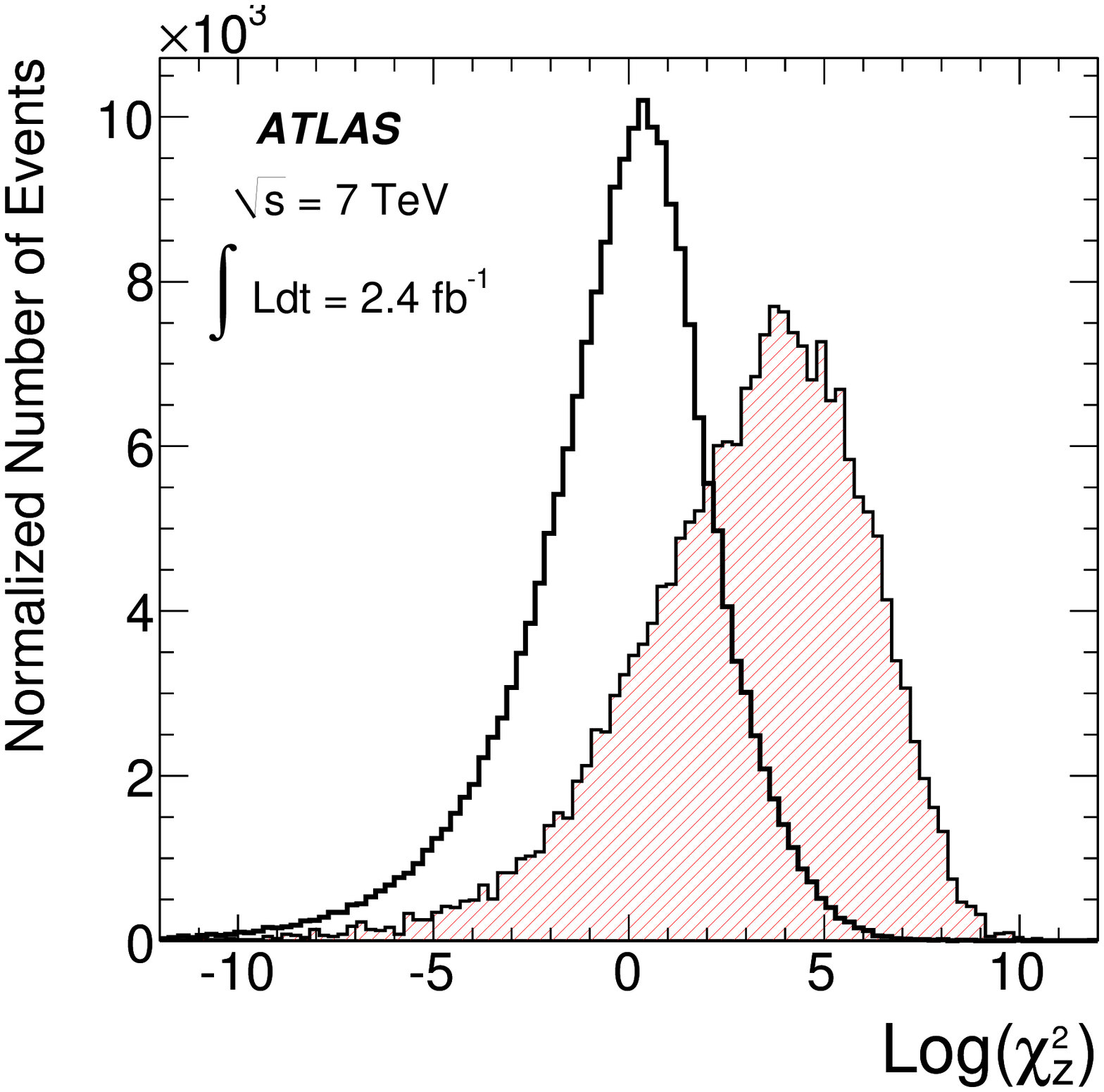}
\includegraphics[width=0.25\linewidth]{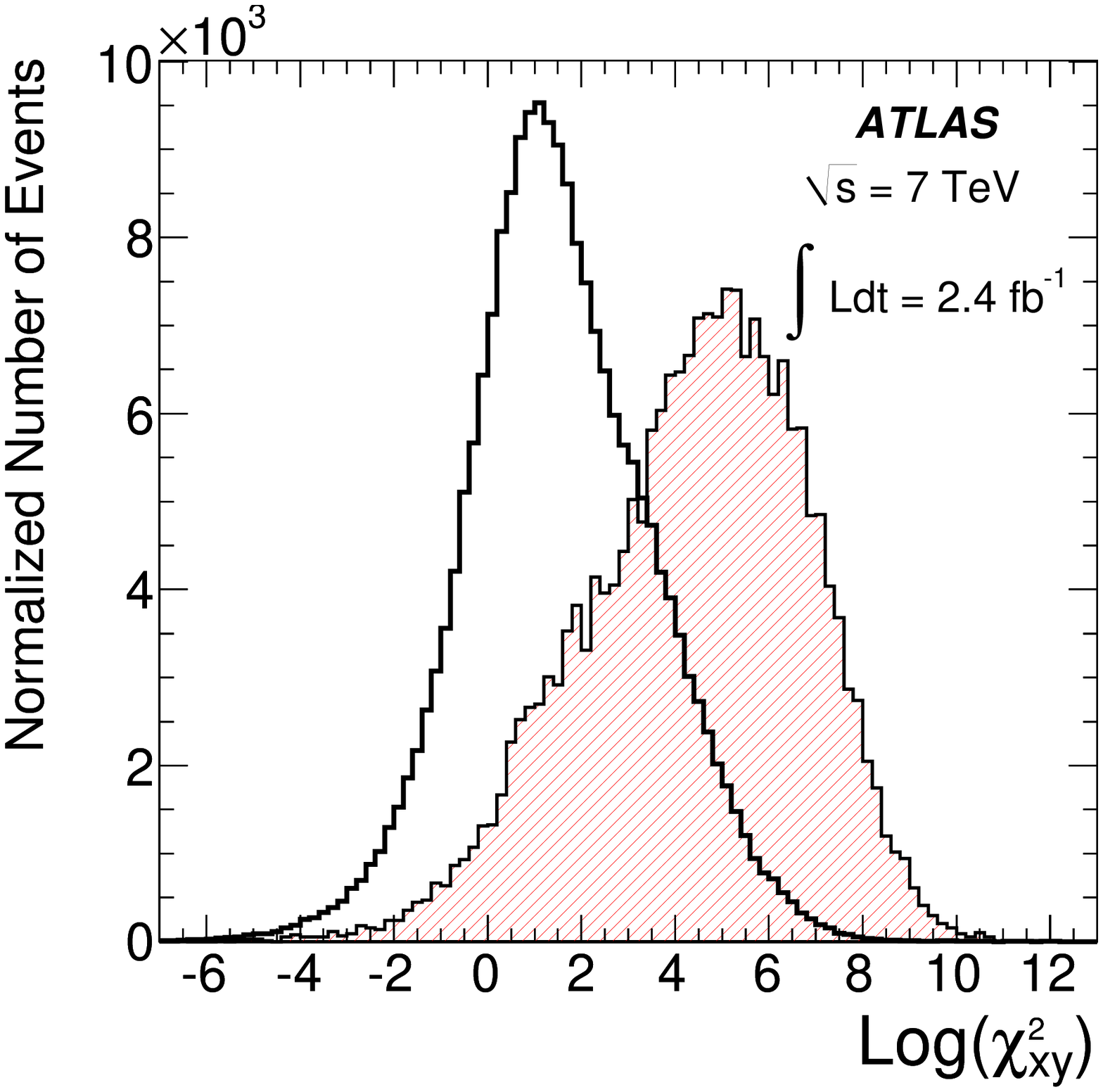} \\
\includegraphics[width=0.25\linewidth]{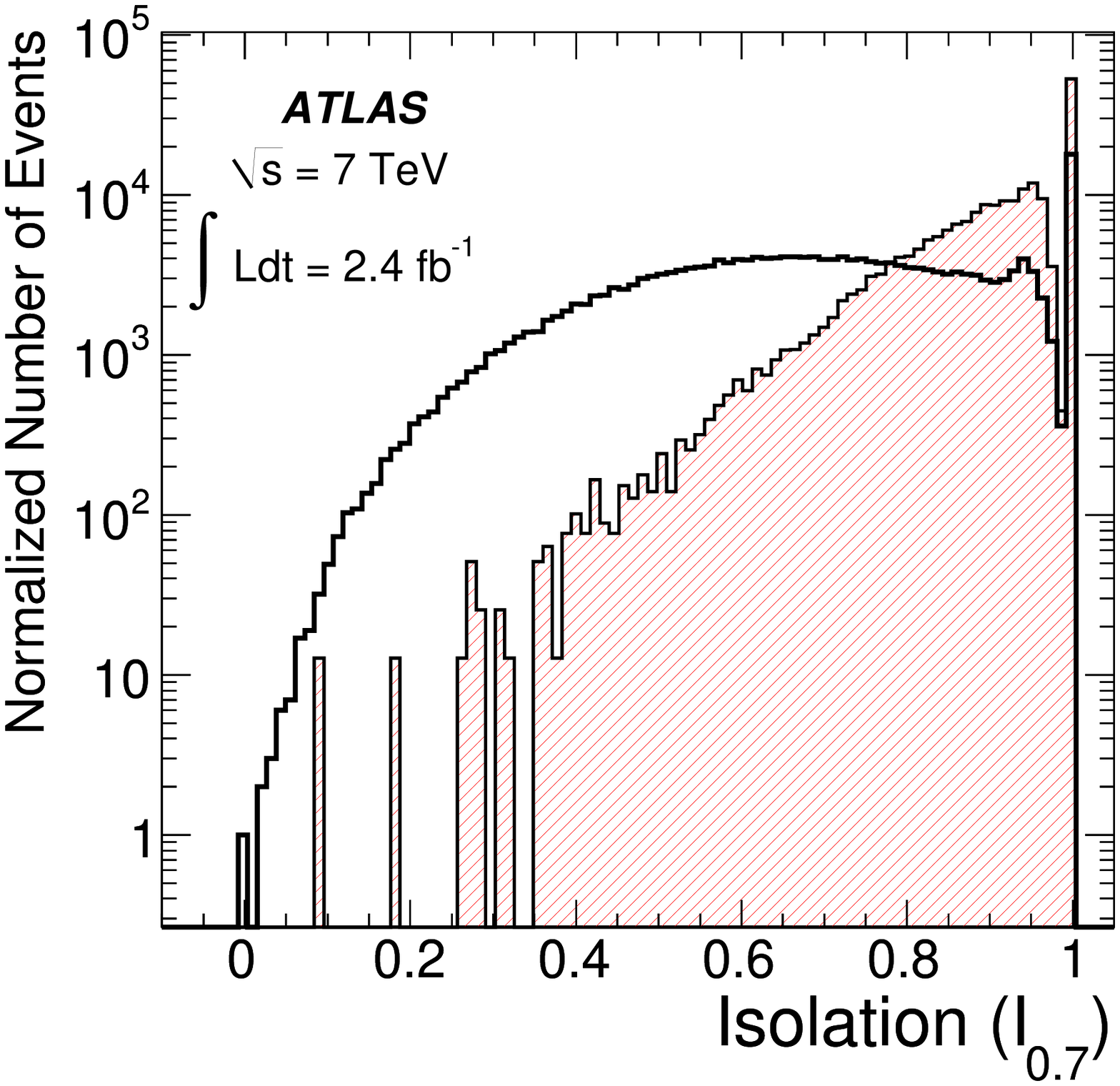}  
\includegraphics[width=0.25\linewidth]{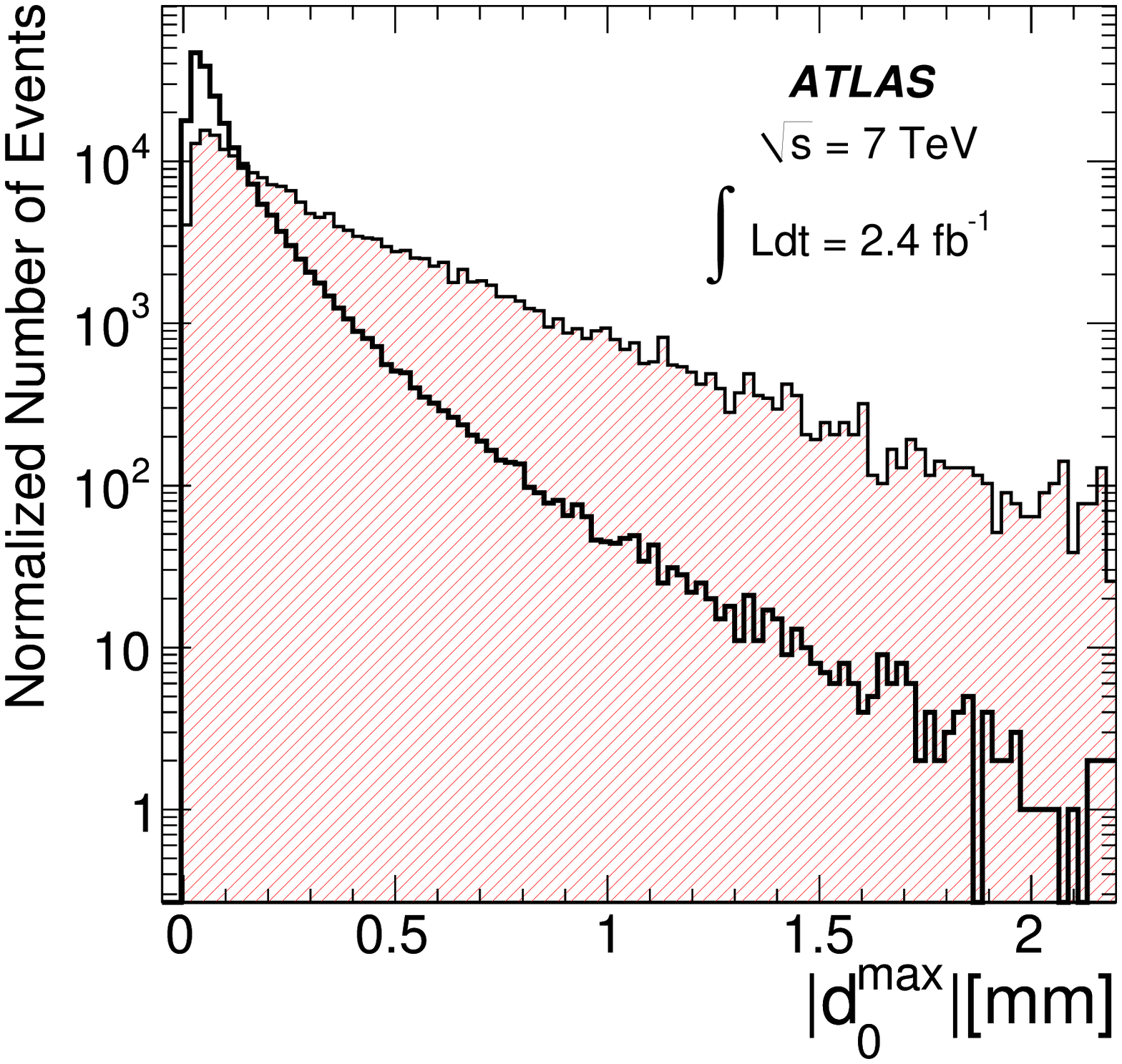}
\includegraphics[width=0.25\linewidth]{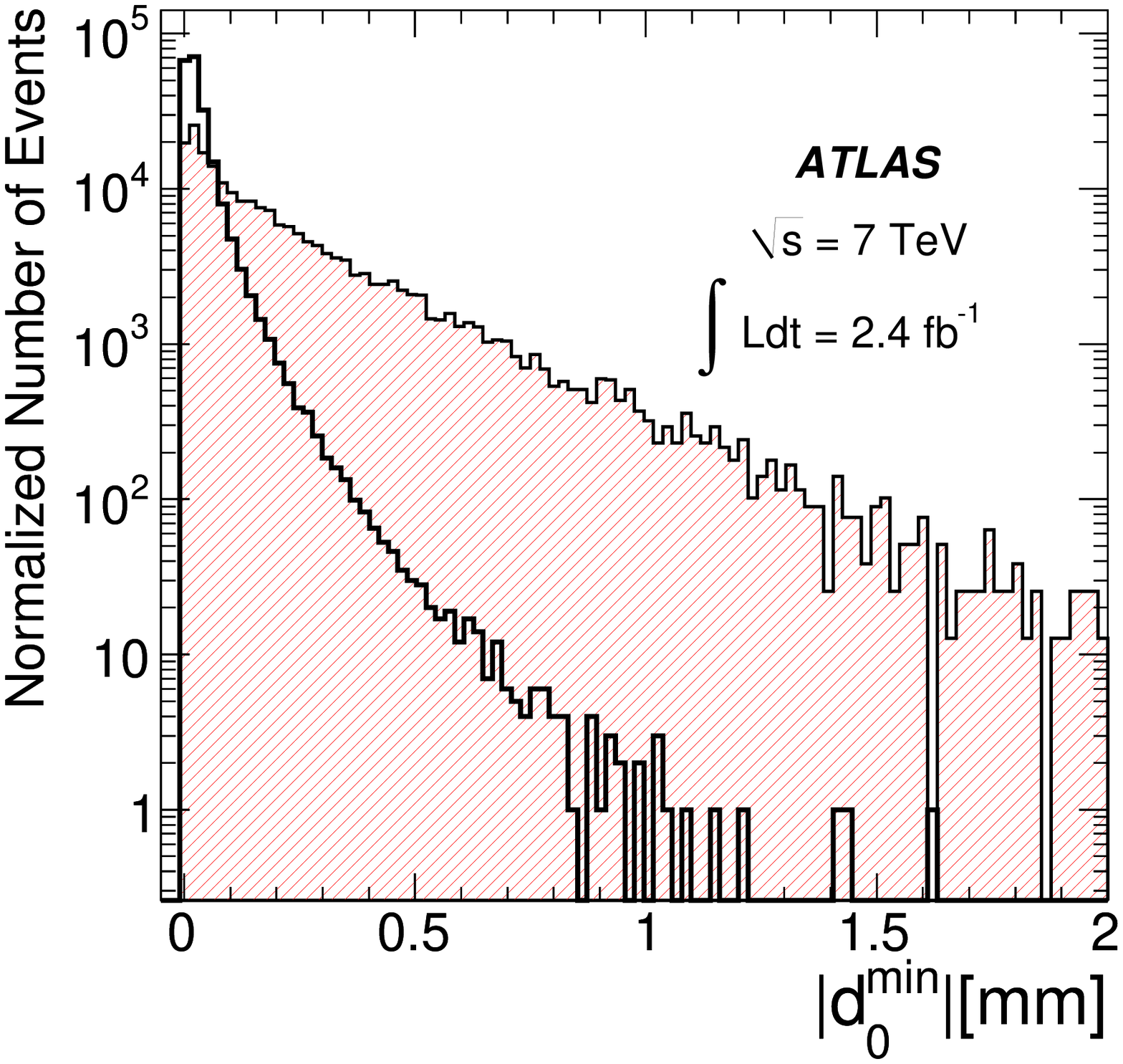} \\ 
\includegraphics[width=0.25\linewidth]{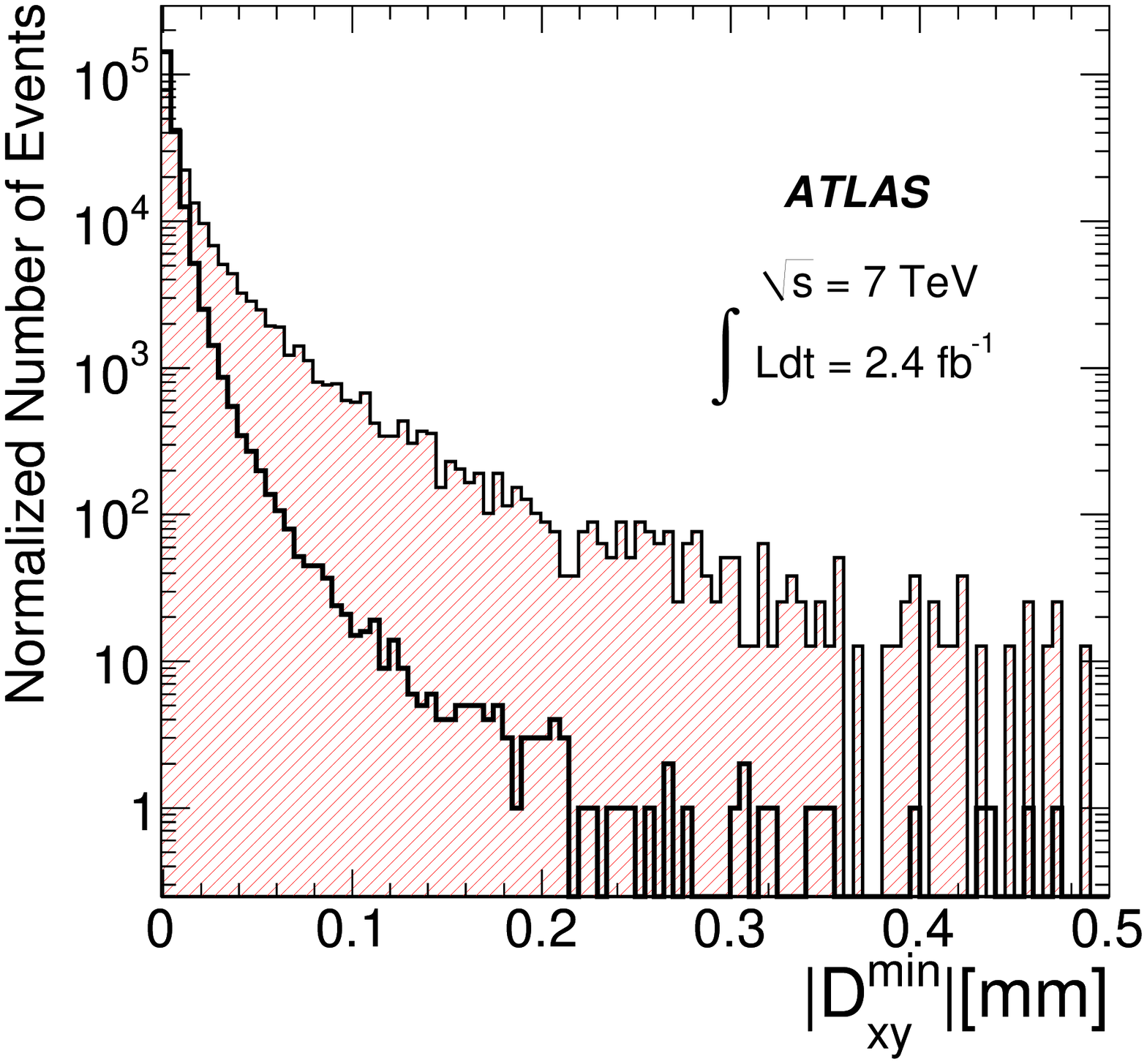}
\includegraphics[width=0.25\linewidth]{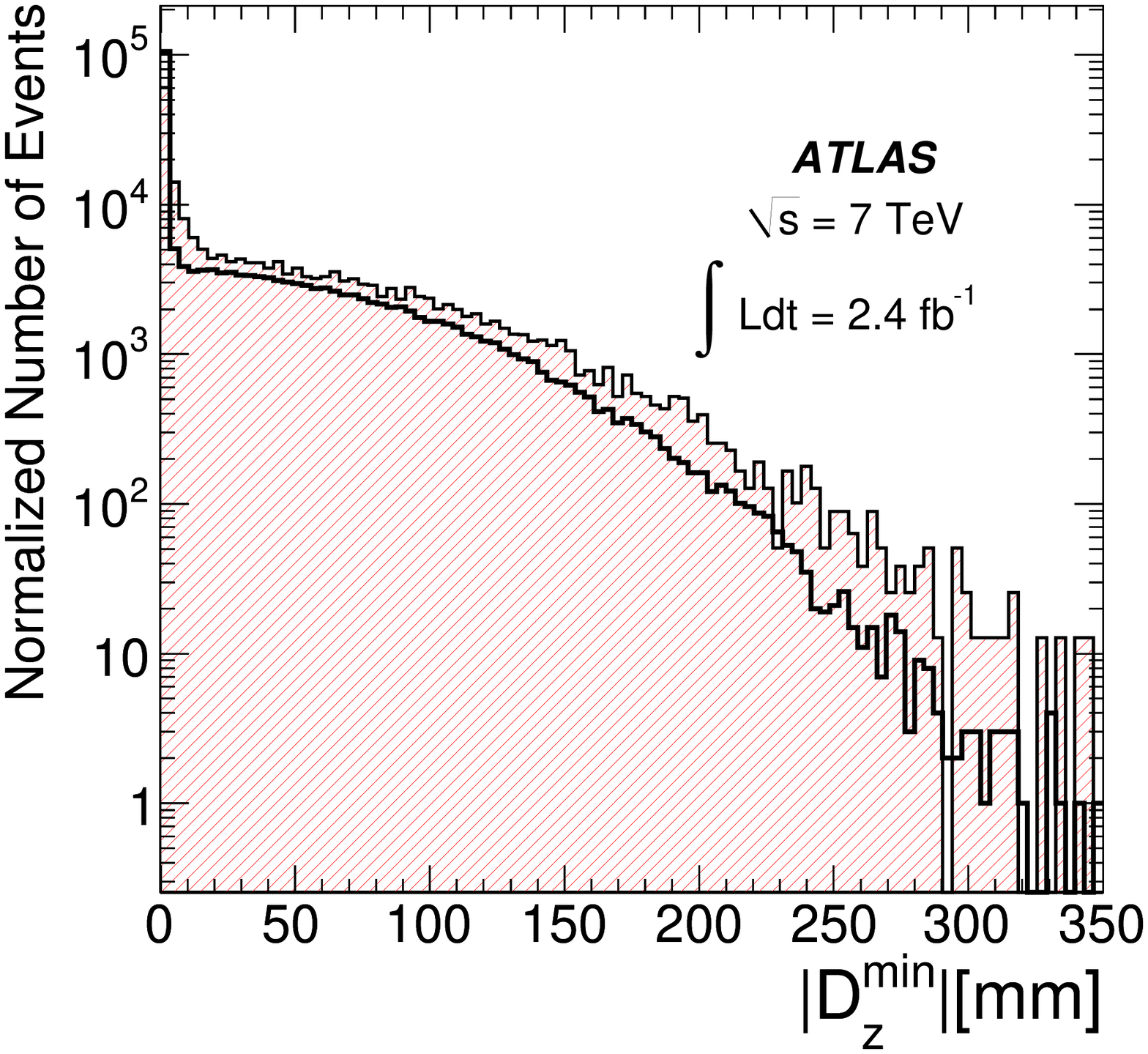}  
\includegraphics[width=0.25\linewidth]{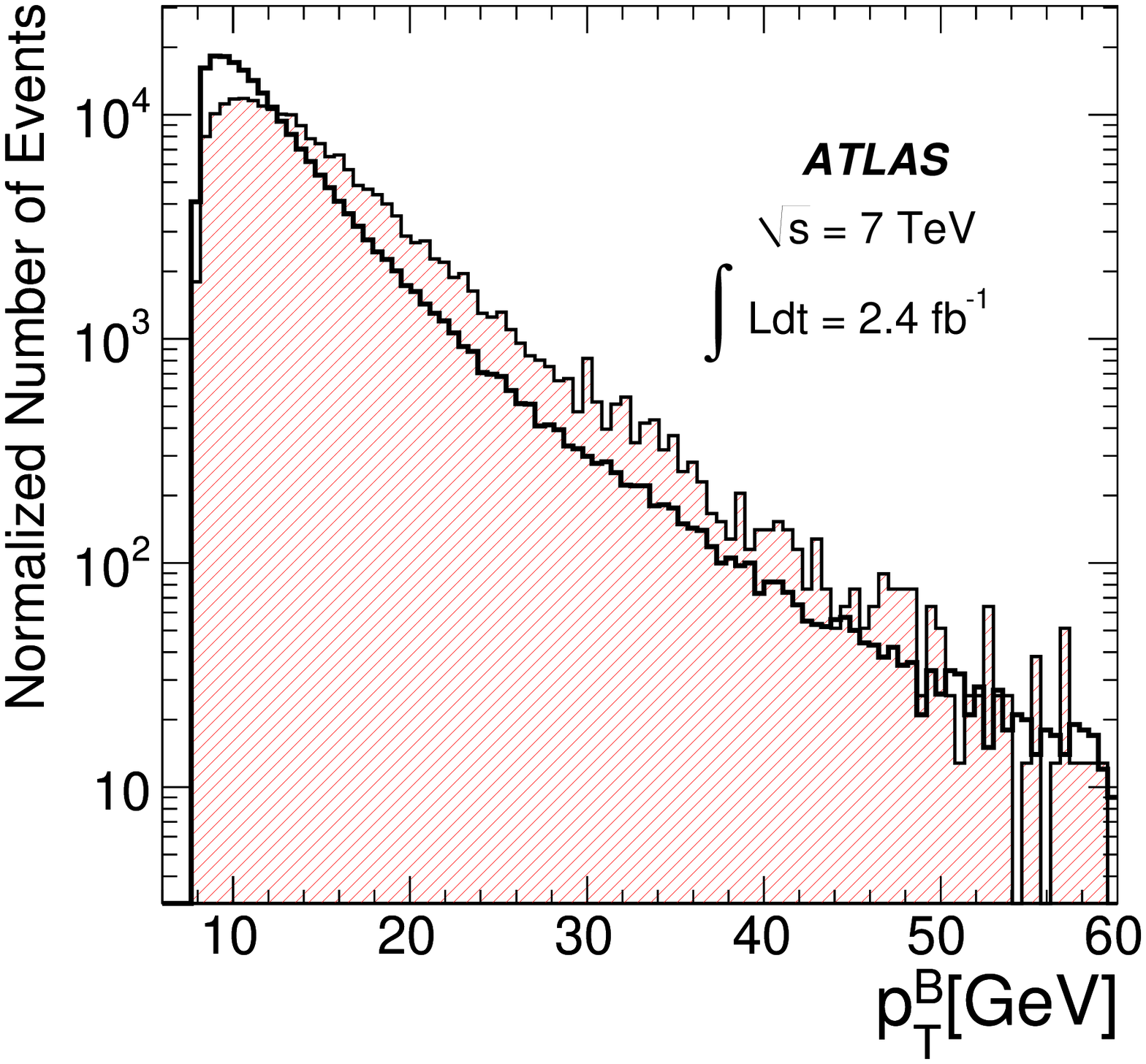}  \\
\includegraphics[width=0.25\linewidth]{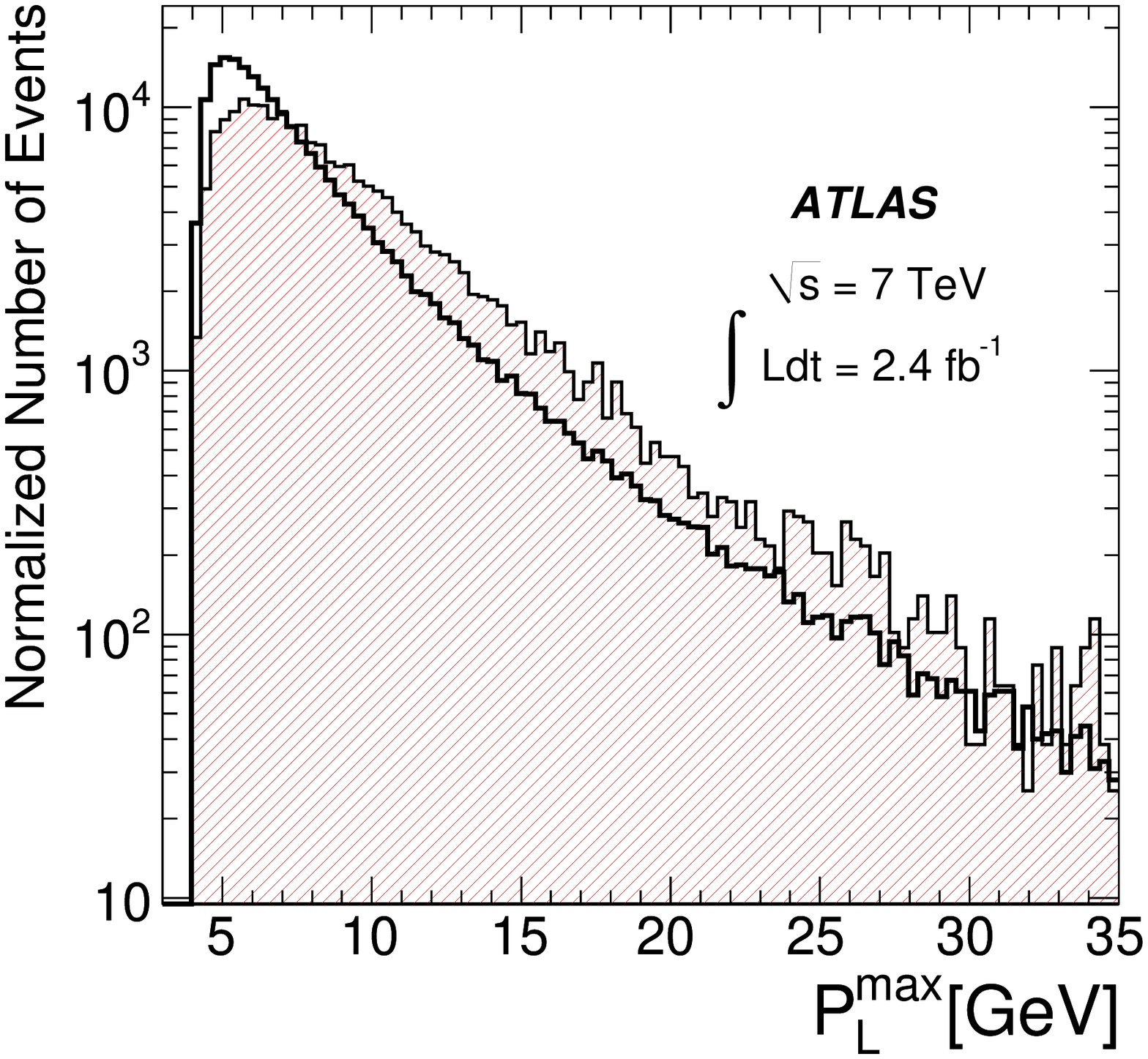}  
\includegraphics[width=0.25\linewidth]{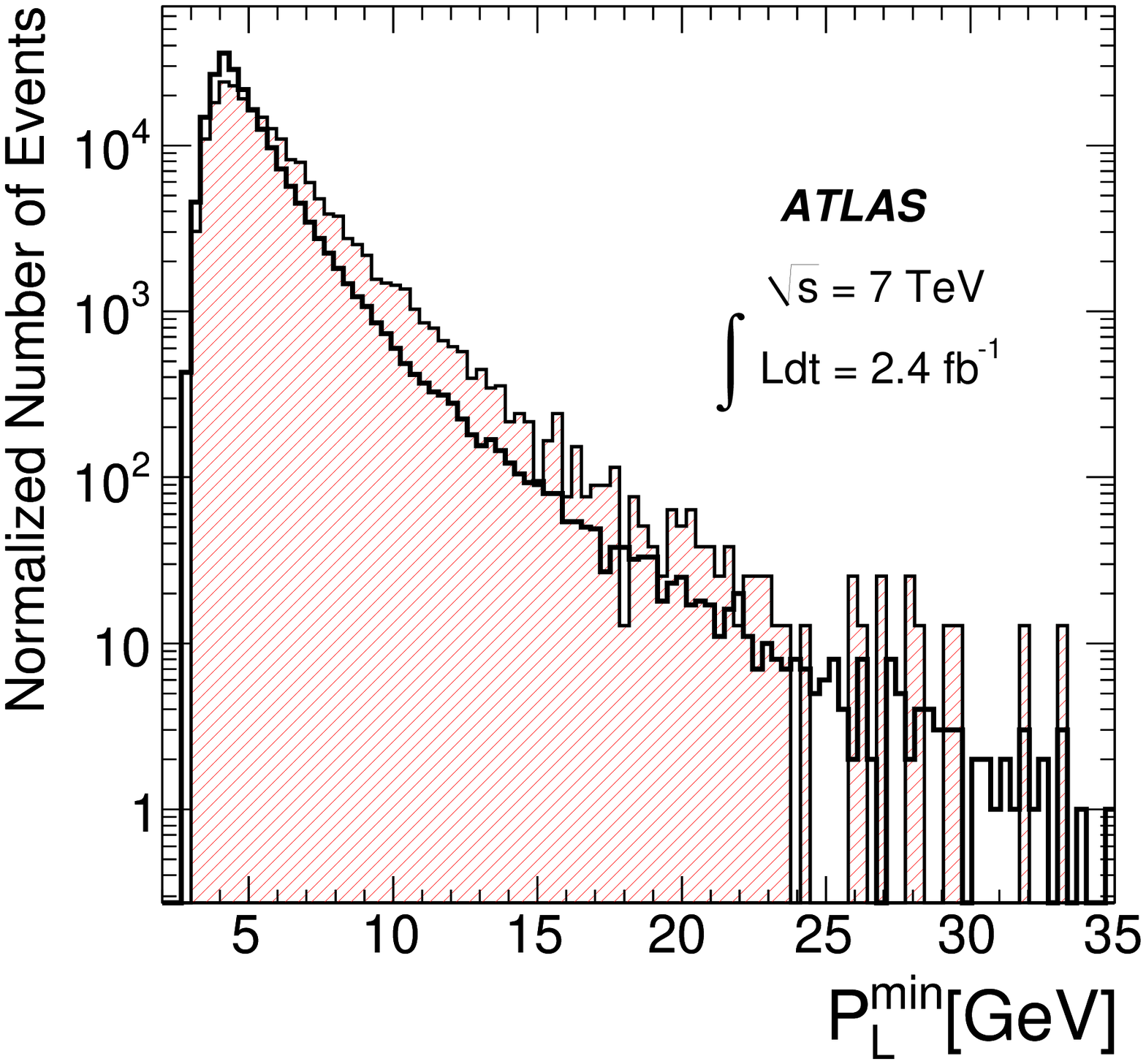}  
\caption{Signal (filled histogram) and sideband (empty histogram) distributions for the selection variables described in Table \ref{table:discri-var}. The $\Bs \to \mumu$ signal (normalized to the background histogram) is from simulation and the background is from data in the invariant-mass sidebands.}
\label{fig:separation_variables}
\end{figure*}

Figure~\ref{fig:separation_variables} shows how the discriminating variables are distributed for signal and background.
Among the discriminating variables, isolation ($I_{0.7}$) is expected to have the largest pile-up dependence. In order to minimize this dependence, 
the definition of $I_{0.7}$ was restricted to only include tracks originating from the primary vertex associated with the $B$ decay.
\begin{figure}[htb!] 
\begin{center}
\includegraphics[width=0.8\linewidth]{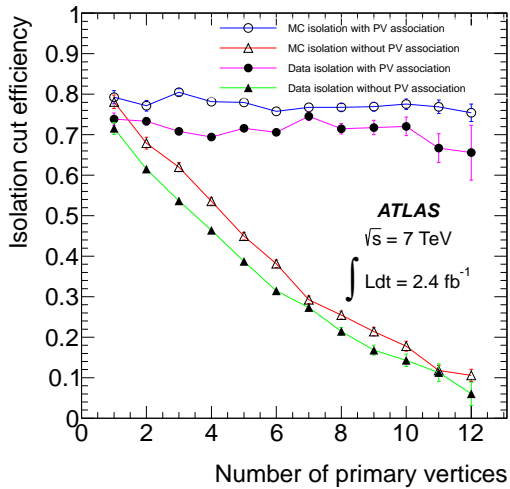}
\caption{Efficiency of the cut $I_{0.7}>0.83$ as a function of
the primary vertex multiplicity for \BpmKpmJpsi\  candidate events from data (filled symbols) and MC simulation (empty symbols). The triangles show the efficiency when including all the tracks in the event, while
circles show the same efficiency with the isolation definition used in this analysis.} 
\label{fig:IsolationPerformanceWithPileup}
\end{center}
\end{figure}
This specification makes the selection independent of pile-up, as shown in 
Figure~\ref{fig:IsolationPerformanceWithPileup}, where the efficiency of the
selection for $B^\pm \to J/\psi K^\pm$ is shown for events with different numbers of reconstructed primary vertices, both in sideband-subtracted data and MC.

The variable $I_{0.7}$ might also be subject to differences between \Bs\ and $B^\pm$ in the distributions of the surrounding hadrons, e.g. with harder $p_T$ spectra for kaons produced in association with the $B^0_s$ in the b-quark fragmentation. 
As predicted by MC, significant differences were observed between \BpmKpmJpsi\ and the control channel \BsJpsiPhi\ in the $I_{0.7}$ distribution from data.
Within statistical uncertainties, the $I_{0.7}$ distribution from the MC simulation of the control channel \BsJpsiPhi\ was verified to be consistent with the corresponding sideband-subtracted signal in data.

\subsection{MC re-weighting and comparison to data}  \label{subsec:MCtuning}
Monte Carlo samples were produced for the signal, reference and control channels, with specific requirements on the $B$ meson decay products as described above in Section \ref{sec:det-prel}.
In order to ensure that the data are reproduced as closely as possible,
the simulation was tuned by an iterative re-weighting procedure: a generator-level (GL) re-weighting based on simulation, followed by a data driven (DD) re-weighting.

For the GL re-weighting, additional MC samples were generated without selection on the final states and 
over a wider range in the $b$-quark kinematics: $\left|\eta^b\right| < 4$ and $p_{\mathrm T}^b > 2.5$~\GeV.
These samples allowed a binned $\left(\pt^B,\eta^B\right)$ map of the efficiencies of the generator-level selections to be derived for both the signal and the reference MC. 
The inverse of such efficiencies was then used to weight events individually, thus correcting the GL biases.
These corrections were applied independently to the simulated reference and signal channel samples
Ê to correct for the biases in the relative $B^0_s/B^\pm$ acceptance induced by the generator-level selection. Possible residual biases were found to be negligible
within the fiducial region $\left|\eta^B\right|<2.5$ and $p_{\rm T}^B>8.0$~\GeV.

Residual $\left(p_{\rm T}^B,\eta^B\right)$ differences between data and MC were observed after GL reweighting. 
These were addressed with the DD re-weighting procedure, based on
the comparison of MC events to the large sample of $B^\pm\to J/\psi K^\pm$ decays in collision data.  
In order not to correlate the re-weighting procedure with the yield measurement, only candidates with odd event numbers in the ATLAS dataset were used in this procedure, while the remaining sample was used for the yield measurement.  

DD weights were determined by an iterative method, comparing re-weighted MC events with sideband-subtracted
 $B^\pm\to J/\psi K^\pm$ events in data. The procedure was applied separately to the $B$ meson variables $p_{\rm T}^B$ and $\eta^B$ due to the limited number of reconstructed $B^\pm\to J/\psi K^\pm$ events in data, deriving the weights:
\begin{equation}
W_{ij}(p_{\rm T}^B,\eta^B)=w_i(p_{\rm T}^B) \times  w_j(\eta^B)\label{formula:Weights}
\end{equation}
where $W$ represents the final DD weights,
 the indices $i$ and $j$ refer to bins in $p_{\rm T}^B$ and $\eta^B$, and $w_k=N_k^{data} /N_k^{MC}$  
is the data-to-MC ratio of the normalized number of entries for each variable.
The convergence and the consistency of the procedure, together with the factorization assumption of Eq.~\ref{formula:Weights}, were tested with additional MC samples, where intentionally distorted $(\pt^B,\eta^B)$ spectra were found to converge to the expected distributions.
Effects related to the finite resolution in the measured variables were estimated to be smaller than $1\permil$ of the bin content and are therefore negligible when compared to statistical uncertainties.  

Generator-level biases were addressed by applying the GL re-weighting before the DD re-weighting, and by verifying that this correction yields compatible $(\pt^B,\eta^B)$ spectra for $\Bs\to\mumu$ and $B^\pm\to J/\psi K^\pm$ MC samples.
Finally, the full re-weighting procedure was applied to \BsJpsiPhi\ decays,
verifying within statistical uncertainty the consistency of the weights with those from $B^\pm\to J/\psi K^\pm$.

\begin{figure}[htb!]
\begin{center}
\subfigure[]{
\includegraphics[width=0.9\linewidth]{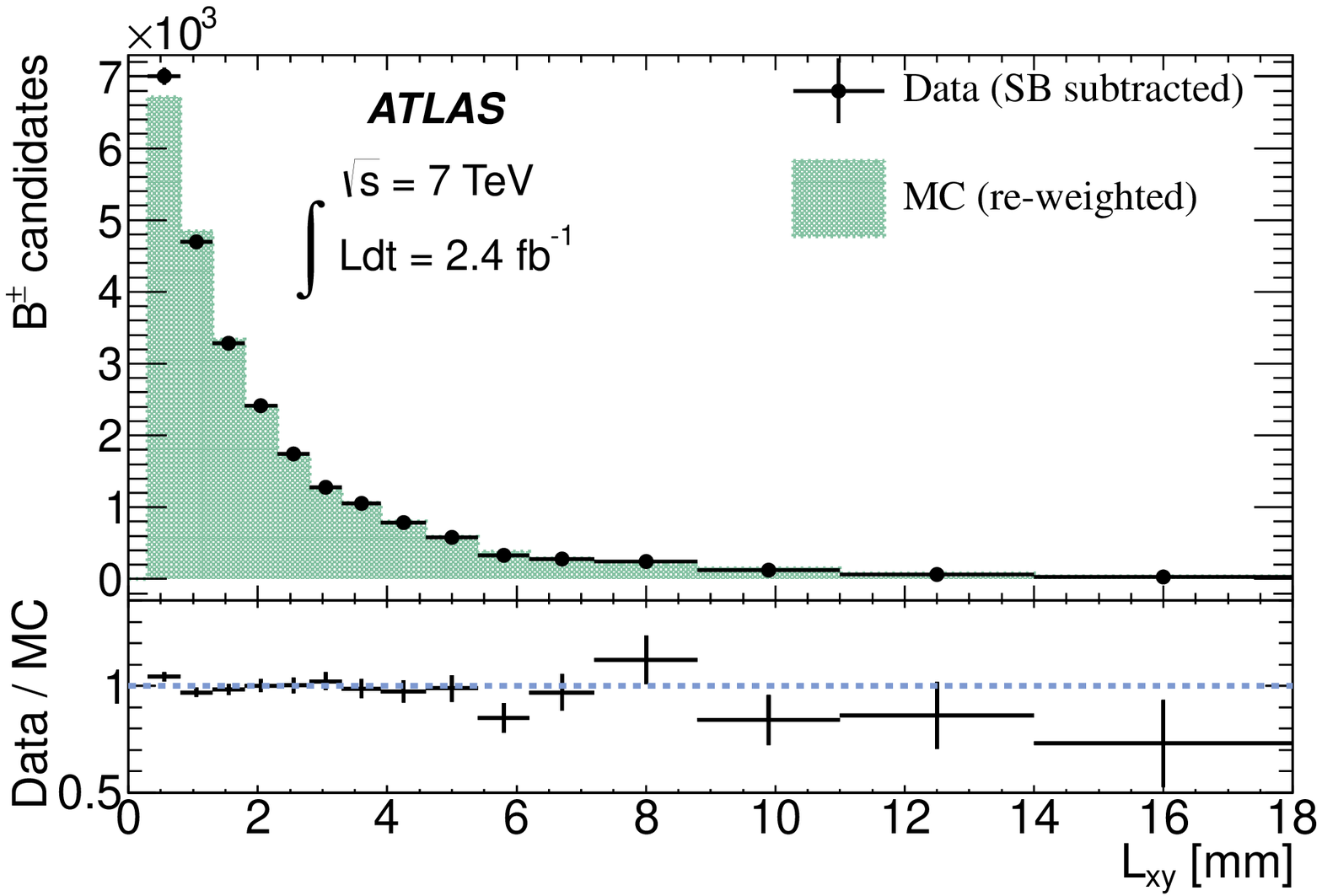}
}
\subfigure[]{
\includegraphics[width=0.9\linewidth]{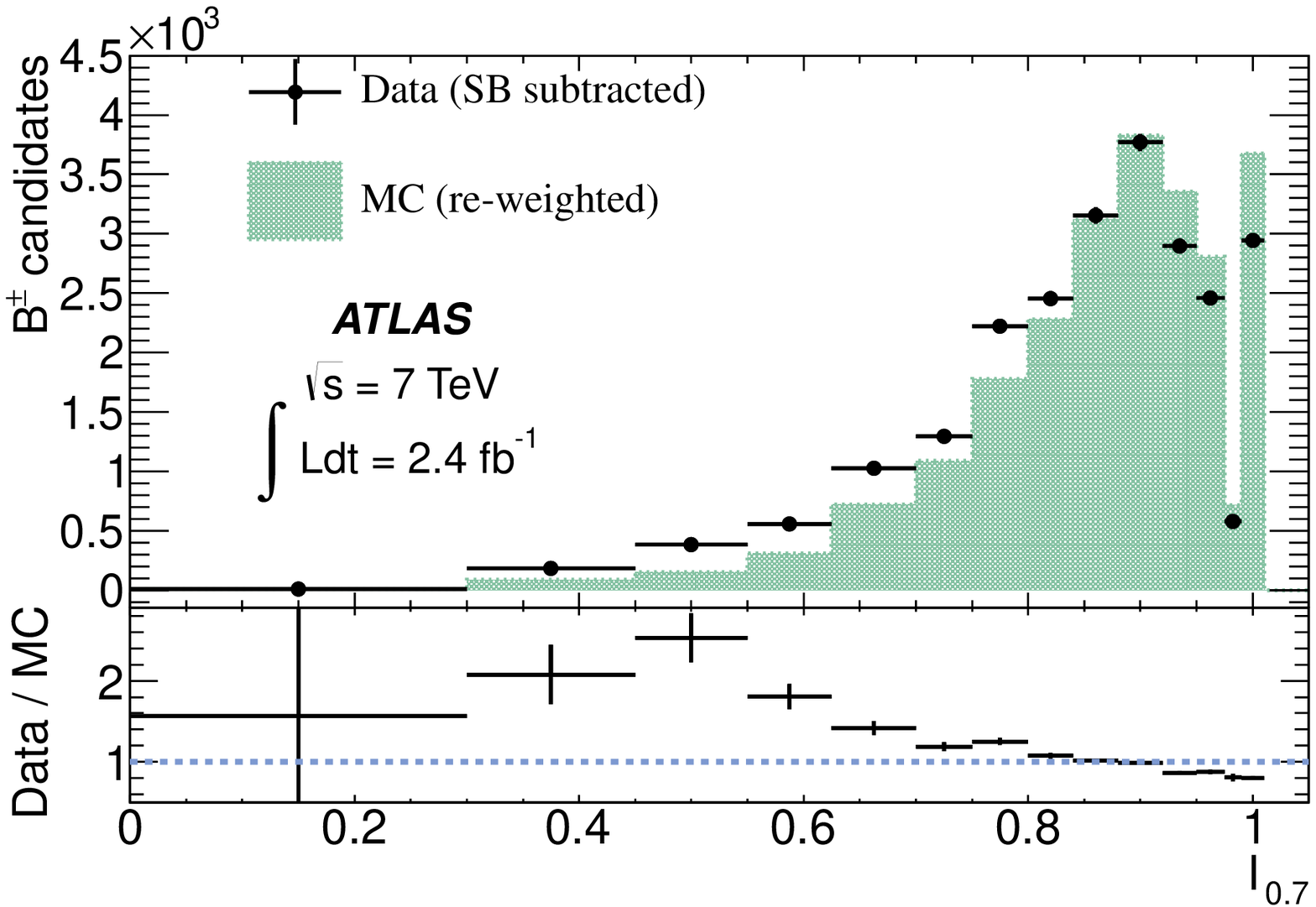}
}
\caption{\label{fig:main:DataMCComparisons}Examples of sideband-subtracted data-reweighted MC comparisons using $B^\pm\to J/\psi K^\pm$ decays for two of the most powerful separation variables: (a) $L_{\rm xy}$ and (b) $I_{0.7}$. Uncertainties are statistical only.
The lower graph in each case shows the data/MC ratio.}
\end{center}
\end{figure}

Distributions from $B^\pm\to J/\psi K^\pm$ in MC simulation and data  were compared,
after side-band background subtraction, for all discriminating variables listed in Table~\ref{table:discri-var} and for variables used in the preselection.
Agreement between MC and data was found for most of the variables. Figure~\ref{fig:main:DataMCComparisons} shows comparisons for $L_{\rm xy}$ and $I_{0.7}$. 
Systematic effects associated with the residual data-MC differences are discussed in Section~\ref{sec:reference-channel}.
The uncertainties on the GL$\times$DD weights are dominated by systematic uncertainties obtained from the comparison between data and MC. They were propagated through the analysis and included among the systematic uncertainties in the signal extraction, as discussed in Section~\ref{sec:extraction}. 

\subsection{Selection optimization} \label{subsec:TMVA-1}
The optimization of the event selection was performed by maximizing the estimator:
\begin{equation}
\mathcal{P}=\frac{\epsilon_\mathrm{sig}}{\frac{a}{2}+\sqrt{N_\mathrm {bkg}}}\, ,
\label{formula:Punzi}
\end{equation}
where $\epsilon_\mathrm{sig}=\mathcal{A}_{\mu^+\mu^-}\epsilon_{\mu^+\mu^-}$ and $N_\mathrm {bkg}$ are the signal acceptance times efficiency relative to the simulated phase space of the samples in section \ref{subsec:MCtuning} (corresponding to the signal efficiency defined for $|\eta^B|<2.5$ and $\pt^B>8.0$~\GeV) and the background yield for a given set of cuts. The extraction of $N_\mathrm{bkg}$ is performed by sideband interpolation as described in section \ref{sec:extraction}. The coefficient $a$ was determined by the 
confidence level (CL) sought in the analysis, with $a$=2 for a $95\%$ CL limit. This quantity is specifically designed to optimize the performance of a frequentist limit determination
in a counting analysis \cite{PunziPaper}. 

First, a simplified optimization procedure was performed on a small set of variables that includes:
$|\alpha_{2\mathrm D}|$, $I_{0.7}$, $ct$, and width $\pm\Delta m$ of the search 
window centred around the \Bs\ mass (rounded to $5366$~\MeV). A four-dimensional scan was performed on the four variables, using odd-numbered events in the sidebands. 
The optimal selection cuts are shown in Table~\ref{tab:main:CutScanResults}, where  the signal efficiency $\mathcal{A}_{\mu^+\mu^-}\epsilon_{\mu^+\mu^-}$, the background estimated from sidebands interpolation and the value of $\cal P$ are also given. This selection serves
as a benchmark for the optimization of the multivariate analysis described in Section \ref{subsec:TMVA-2}.



\begin{table}[h]
\begin{center}
\setlength{\tabcolsep}{4pt}
\begin{tabular}[h]{cccc|ccc}
\hline\hline
$|\alpha_{2\mathrm D}|$ & $ct$ &  $I_{0.7}$ & $\Delta m$ &  $\epsilon_{\rm sig}$ &  $N_\mathrm {bkg}$ &  $\mathcal{P}$ \\
 $<$0.03 & $>$0.3 mm & $>$0.83 & $\pm$ 105 \MeV & 0.040 & 9$\pm$2 &  0.010 \\
\hline\hline
\end{tabular}
\caption{Optimal selection variable cuts for the four-variables scan, and resulting analysis performance
in terms of signal acceptance times efficiency ($\epsilon_{\rm sig}$), background yield in the signal region ($N_{bkg}$) and the estimator $\mathcal{P}$.}
\label{tab:main:CutScanResults}
\end{center}
\end{table}

\subsubsection{Categories of invariant-mass resolution} \label{subsec:TMVA-3-res-families}
The ability to resolve a small $\Bs \to \mumu$ signal from the continuum background depends on the width
$\Delta m$ of the search region and is therefore affected by the resolution.
The latter varies considerably over different sub-samples of muon pairs measured by ATLAS, due to the increase in multiple scattering and the decrease of the magnetic field integral at large values of $|\eta|$.
The non-resonant background invariant mass distribution was observed to be relatively independent of $\eta$.
As a consequence, different mass-resolution categories correspond to different signal-to-background conditions.

In the statistical analysis, regions of different mass resolution and hence signal-to-background ratio were separated in order to optimize them independently.
The sample was separated into three categories, defined by the larger pseudorapidity value $|\eta|_{\mathrm max}$ of the two muons in each event. The three categories were defined by the intervals 
$|\eta|_{\mathrm max}=$  0--1, 1--1.5 and 1.5--2.5.
The corresponding average values of the mass resolution are approximately 60, 80 and 110~\MeV, respectively. 
The relative population of each interval, in $\Bsmumu$ signal MC, amounts to 51\%, 24\% and  25\%.

The same classification, based on $|\eta|_{\rm max}$, was used for the reference channel
$B^\pm\to J/\psi K^\pm$, and separate values of the acceptance-times-efficiency ratio were obtained for each category, as discussed in Section~\ref{subsec:rho}.

\subsubsection{Multivariate selection} \label{subsec:TMVA-2}
The selection with optimal cuts was used to validate the multivariate analysis tool used for the final results.
The TMVA package \cite{TMVA} implementation of Boosted Decision Trees (BDT) was found
to have the best performance and was selected for this analysis.
As a first step, it was verified that for fixed values of $\Delta m$, the optimal BDT corresponds to selections in the variables $\alpha_{2\mathrm D}$, $ct\,$ and $I_{0.7}$ directly comparable to those obtained with the cuts shown in Table ~\ref{tab:main:CutScanResults}.
 Next, the discriminating variables of Table \ref{table:discri-var} were introduced one-by-one into the BDT, 
verifying that the multivariate optimization increased the signal efficiency and the value of $\mathcal{P}$.
With the BDT approach the ${\cal P}$ estimator improved from $\mathcal{P}=0.010$ found in the simplified optimization to $\mathcal{P}=0.016$.

\begin{figure}[htb!] 
\center
\includegraphics[width=\linewidth]{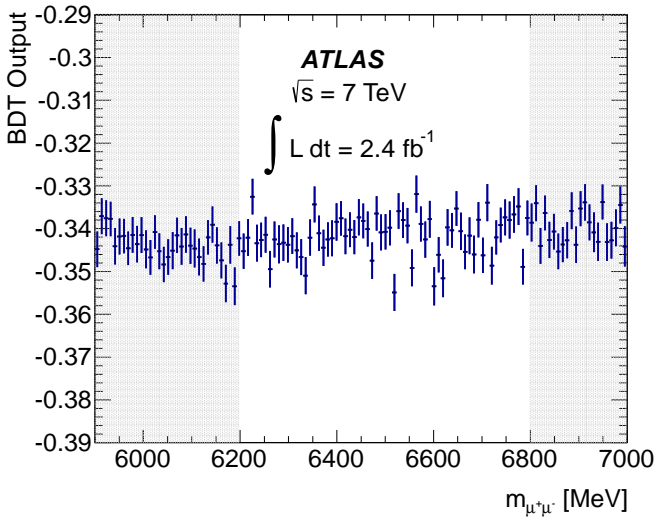}
\caption{Mean and RMS (error bars) of the BDT output in bins of di-muon invariant mass, for background events in the region 5900 to 7000~\MeV, with the 6200 to 6800~\MeV\ region not used in the training of the classifier.
The BDT used is the one trained for the search of the fictitious 6500~\MeV\ signal.}
\label{fig:QvsM}
\end{figure}

In order to avoid biases in the background interpolation, the BDT selection should be insensitive to the mass of the muon pair. 
The BDT inputs have no correlation with the invariant mass.
Residual correlations in the BDT output were studied through the search for a fictitious decay $X\to \mumu$ with $m_X=6500$~\MeV. A Monte Carlo sample was used to provide reference signal events, while data in the mass intervals 5900 to 6200 \MeV\ and 6800 to 7000~\MeV\ were used as background. The BDT training and selection optimization were consistently performed on odd-numbered events. 
Figure~\ref{fig:QvsM} shows the BDT output as a function of the di-muon mass, over the sideband regions and the fictitious signal region (6200 to 6800 \MeV), which was not used in the optimization. No significant mass dependence was observed.

The optimization of the multivariate analysis was performed in the six-dimensional space of $\Delta m$ and the BDT output cuts for each of the mass-resolution  categories.
The independence of the BDT output on $m_{\mumu}$ and the complementarity of the samples allow the factorization of the individual cut efficiencies.
Each efficiency curve was interpolated with analytical models, allowing the numerical maximization of 
$\mathcal{P}$ and yielding the optimal cuts reported in Table \ref{table:6DOptimization}.

\begin{table}[h!]
\begin{center}
\begin{tabular}[h]{l|ccc} 
\hline
$|\eta|_{\rm max}$ Range                                       &        0--1.0         &          1.0--1.5        &              1.5--2.5  \\
\hline\hline
invariant mass window [\MeV]                      &   $\pm 116$  & $\pm 133$  & $\pm 171$\\
BDT output threshold                                     &   0.234           &   0.245          &     0.270     \\
\hline\hline
\end{tabular}
\caption{BDT output and $\Delta m$ cuts for each mass-resolution category, optimized according to the method described in the text.}
\label{table:6DOptimization}
\end{center}
\end{table}

\section{Single Event Sensitivity ingredients} \label{sec:reference-channel}
\subsection{Relative acceptance and efficiency}  \label{subsec:rho}
The ratio of the acceptance times efficiency products for the charged and neutral decays 
\[
\earatio =(A_{J\!/\!\psi K}\,\epsilon_{J\!/\!\psi K})/(A_{\mumu}\,\epsilon_{\mumu})
\]
 is required for the determination of the SES (Eq.~\ref{BRFormula}).
The same BDT, trained on the $\Bs$ signal MC sample and di-muon data sidebands, was used to select both decay modes.

The uncertainty on $\earatio$ is affected by differences between data and MC in the distributions of the discriminating variables. 
Such differences are reduced by the data-driven corrections applied to the MC $B$-meson kinematics.
Furthermore, only deviations that act incoherently between the signal and the reference channel contribute to the uncertainty on $\earatio$.
These effects were studied by observing the change in the relative efficiency of the BDT selection 
when the simulated events were re-weighted by the data-to-MC ratio of the distributions of the most
sensitive variables in $B^\pm\to J/\psi K^\pm$ events.
The procedure was performed with the cut on the BDT output fixed at the optimal value for each of the three event categories.
Conservatively, the corresponding variations in $\earatio$ were combined linearly and taken as systematic
uncertainties.

Due to large correlations between $L_{\rm xy}$, $\chi^2_{\rm xy}$ and $ct$-significance, 
correcting for the differences in $L_{\rm xy}$ between data and simulation was found to also effectively remove differences in the other two variables. 
Therefore only $L_{\rm xy}$ was considered, since it induced the largest deviation in $\earatio$.
Differences in the $\eta$ and $p_{\mathrm T}$ distributions of the final state particles, the hit multiplicity in the Pixel detector, and the multiplicity of reconstructed primary vertices were included in the systematic uncertainty evaluation.

\begin{figure}[hbt!] 
\center
\includegraphics[width=0.8\linewidth]{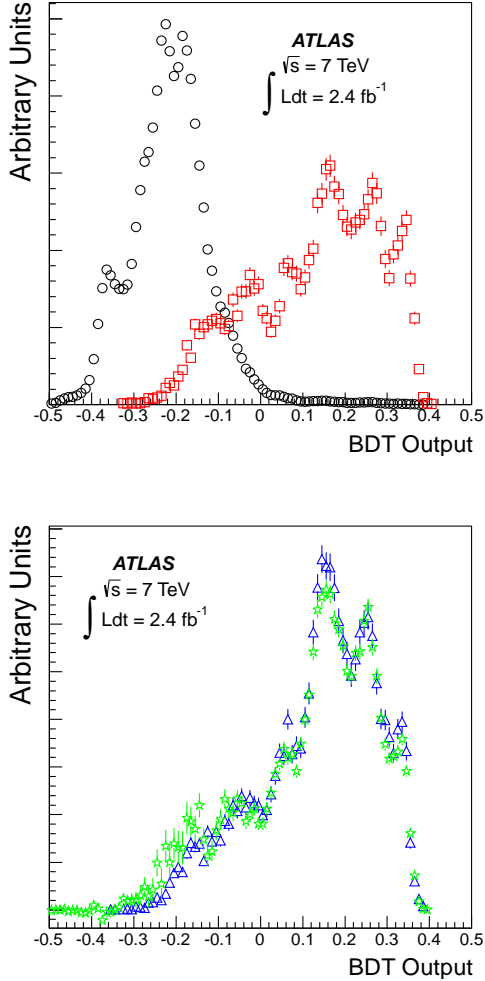}
\caption{Distributions of the response of the BDT classifier. Top: $\Bs \to\mumu$ MC sample (squares) and data sidebands (circles);
bottom: $\BpmKpmJpsi$ events from tuned MC samples (triangles) and sideband-subtracted data (stars). }\label{fig:main:qResponseVariation} 
\end{figure}

Figure \ref{fig:main:qResponseVariation} shows the distribution of the BDT output for MC samples of $B_s^0 \to \mumu$ and $B^\pm\to J/\psi K^\pm$ decays, with a signal--background
comparison for $\Bsmumu$ and a sideband-subtracted data--MC comparison for $B^\pm\to J/\psi K^\pm$.
As shown in Table \ref{table:6DOptimization}, the selection required the BDT output to exceed 0.23--0.27, depending on the mass-resolution category. 
The systematic uncertainties induce a fractional change in the number of events passing the BDT cut varying between 10\% and 20\% depending on the category. 
This change is highly correlated between the two channels: the corresponding variation on the efficiency ratio is $0.6\%$, which was taken as a systematic uncertainty and is smaller than the $\pm 2.3\%$ error due to the finite MC statistics. 

The value of $\earatio$ and its systematic uncertainties (shown in Table~\ref{tab:RhoFamilies}) were derived separately in the three mass-resolution categories.  
The MC-based efficiency was compared with that from \BpmKpmJpsi\  data, computing the efficiency of the BDT cut relative to the preselection. The results are of similar precision and fully consistent:
$0.258 \pm 0.013\, ({\rm stat})$ for the data and $0.234 \pm 0.014  \, ({\rm stat}) \pm 0.011\, ({\rm syst})$ for MC.

Additional smaller contributions to the uncertainty on $\earatio$ are due to the data-MC discrepancy in vertex reconstruction efficiency ($\pm2\%$) \cite{mc10detail}, the uncertainty on the absolute
$K^\pm$ reconstruction efficiency as derived from simulation of the $B^\pm \to J/\psi K^\pm$ reference channel ($\pm 5\%$) and asymmetry differences in detector response to $K^+$ and $K^-$ mesons ($\pm 1\%$).

 \begin{table}[h!]
\begin{center}
\begin{tabular}[h]{c c c c} 
\hline
  $\left|\eta\right|_{\rm max}$ & $\earatio^i$ & $\Delta$ \%  & $\Delta$ \%  \\
  Range  &        &       Stat.             &    Syst.                 \\

\hline\hline
 0--1.0 & 0.274 & 3.1 & 3.1 \\
 1.0--1.5 & 0.202 & 4.8 & 5.5 \\
 1.5--2.5 & 0.143 & 5.3 & 5.9 \\
\hline\hline
\end{tabular}
\caption{ Values of the acceptance-times-efficiency ratio $\earatio$ between reference and search channel,
shown separately for the different categories in mass-resolution.}
\label{tab:RhoFamilies}
\end{center}
\end{table}

\subsection{$B^\pm \to J/\psi K^\pm$ event yield} \label{subsec:Bplus}
\begin{figure}
	\begin{center}
		\includegraphics[width=\linewidth]{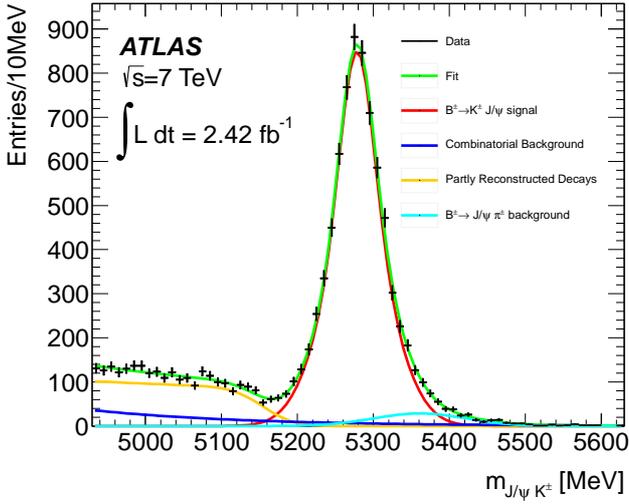}
		\caption{$J/\psi K^{\pm}$ mass distribution for all the $B^\pm$ candidates from even-numbered events passing all the selection cuts, merged for illustration purposes.
		Curves in the plot correspond to the various fit components: two Gaussians with a common mean for the main peak, a single Gaussian with higher mean for the
		$B^\pm \to J/\psi \pi^\pm$ decay, a falling exponential for the continuum background and an exponential function multiplying a complementary error function for
		 the partially reconstructed decays.}
 		\label{fig:bplusfit}
	\end{center}
\end{figure}

The reference channel yield $N_{J/\psi K^\pm}$ was determined from a binned likelihood fit to the invariant mass distribution of the $\mumu K^\pm$ system, performed in the mass range 4930-5630~\MeV. 
 To avoid any bias induced by the DD re-weighting of the MC samples discussed in Section~\ref{subsec:MCtuning}, only even-numbered events were used in the extraction of the $B^\pm \to J/\psi K^\pm$ event yield. 
The $B^\pm$ signal was modelled with two Gaussian distributions of equal mean value. The background was modelled with the sum of: (a) an exponential function for the continuum combinatorial background; (b) an exponential function multiplied by a complementary error function describing the low-mass ($m < 5200$~\MeV) contribution for partially reconstructed decays (such as $B \to J/\psi K^*$, $B \to J/\psi K$(1270) and $B \to \chi_c K$); and (c) a Gaussian function for the background from $B^\pm \rightarrow J/\psi\pi^\pm$.  
 Figure~\ref{fig:bplusfit} shows the invariant mass distribution and the result of the fit for the selected data sample.

All parameters describing the signal and background were determined from the fit, with the exception of the mass and the width of the last component (c), which were obtained from simulation. The fit was performed for each of the three categories of mass resolution.

Systematic uncertainties affecting the extracted reference yield were estimated by varying the fit model: use of different bin sizes (10 or 25 \MeV\ and unbinned), different models for signal and continuum background, inclusion of event-wise di-muon mass resolution.
The resulting $B^\pm$ yields are given with their statistical and systematic uncertainties in Table~\ref{table:Kplus-yield}.

\begin{table}[h!]
\begin{center}
\begin{tabular}[h]{l|ccc} 
\hline
$|\eta|_{\rm max}$ Range                                       &  0--1.0  &   1.0--1.5 &   1.5--2.5  \\
\hline\hline
$B^\pm \to J/\psi K^\pm \to \mumu K^\pm$  & 4300 & 1410 & 1130  \\	
statistical uncertainty                                      &  $\pm 1.6 \%$ &  $\pm 2.8 \%$ &  $\pm 3.0 \%$ \\
systematic uncertainty                                    &  $\pm 2.9 \%$ &  $\pm 7.4 \%$ &  $\pm 14.1 \%$  \\
\hline\hline
\end{tabular}

\caption{Event yield for even-numbered candidates in the reference channel. 
}
\label{table:Kplus-yield}
\end{center}
\end{table}

\section{Inputs to the limit extraction}\label{sec:extraction}
The evaluation of  the SES requires as input the combined branching fraction for the reference channel \BpmKpmJpsi$\to\mumu K^\pm$, which is
$(6.01\pm0.21)\times 10^{-5}$
 \cite{pdg_book}. The relative production rate of $B^0_s$ relative to $B^\pm$ $f_s/f_u$ is $0.267\pm0.021$ \cite{Aaij:2011jp}, 
 assuming $f_u=f_d$ (following Ref. \cite{Asner:2010qj}) and no kinematic dependence of $f_s/f_u$.
The ratio of acceptance-times-efficiency is discussed in Section~\ref{sec:reference-channel} and  
presented in Table~\ref{tab:RhoFamilies}. 
The branching fractions uncertainties, those on $f_u/f_s$, together with those mentioned in the last paragraph of Section \ref{subsec:rho}, were treated coherently in the three categories of mass resolution.

In each mass-resolution category the $\Bs \to\mumu$ signal yield $N_{\mumu}$ was obtained from the number of events observed in the search window, the number of background events in the sidebands, and the small amount of resonant background discussed in Section~\ref{subsec:background}. 
The expected ratio of the background events in the sidebands to those in the search window is described by the parameter $\thetau_i$, which depends on the width of the invariant-mass interval and on the fraction of events from the sidebands used for the interpolation. The former varies according to the mass-resolution category, and the latter is equal to 50\%, corresponding to the even-numbered events in the data collection.  Uncertainties in the mass dependence of the continuum background produced a $\pm 4$\% systematic error in the value of $\thetau_i$, evaluated by studying the 
variation of $\thetau_i$ for different BDT output cuts and background interpolation models. The systematic variation accounts also for additional background components in the low mass sidebands (e.g. partially reconstructed B decays).
This uncertainty was treated coherently in the three mass-resolution categories.  

\begin{table*}[ht!]
\begin{center}
\begin{tabular}[h]{l|ccc} 
\hline
$|\eta|_{\rm max}$ Range                                       &        0--1.0         &          1.0--1.5        &              1.5--2.5  \\
\hline\hline
SES$=(\epsilon\epsilon_i)^{-1}$ [$10^{-8}$]              & 0.71 & 1.6 &  1.4  \\	
$\epsilon =(f_s/f_u)/{\rm BR}(B^\pm \to J/\psi K^\pm \to \mumu K^\pm)$ [$10^{3}$] &  &  $4.45$ $\pm$ 0.38  & \\
$\epsilon_i=N^{B^\pm\to J/\psi K^\pm}_i/\earatio^i$ [$10^{4}$]                                   & $3.14 \pm 0.17 $   & $1.40 \pm 0.15 $ & $1.58 \pm 0.26 $  \\
bkg. scaling factor $\thetau_i$ & 1.29 & 1.14 & 0.88  \\
sideband count $N^{\rm bkg}_{{\rm obs},i}$ (even numbered events) & 5 & 0 & 2 \\
expected resonant bkg.    $N_i^{B\to hh}$     & 0.10 & 0.06 & 0.08  \\
search region count $N^{\rm obs}_{i}$ & 2   & 1 & 0  \\
\hline\hline
\end{tabular}
\caption{Single event sensitivity and event counts in the three mass resolution categories. The second and third lines report
how the SES$=(\epsilon\epsilon_i)^{-1}$  was split between a coefficient common to all bins, and the per-bin component.
The Table does not include the additional common uncertainties corresponding the sources mentioned in the last paragraph of Section \ref{subsec:rho} ($\pm 5.5\%$ in $\earatio^i$) and to the parameterization of the mass dependence of the continuum background ($\pm 4\%$ in $\thetau_i$).
}
\label{table:SES-counts}
\end{center}
\end{table*}

\begin{figure}[htb!]
\begin{center}
\includegraphics[width=0.95\linewidth]{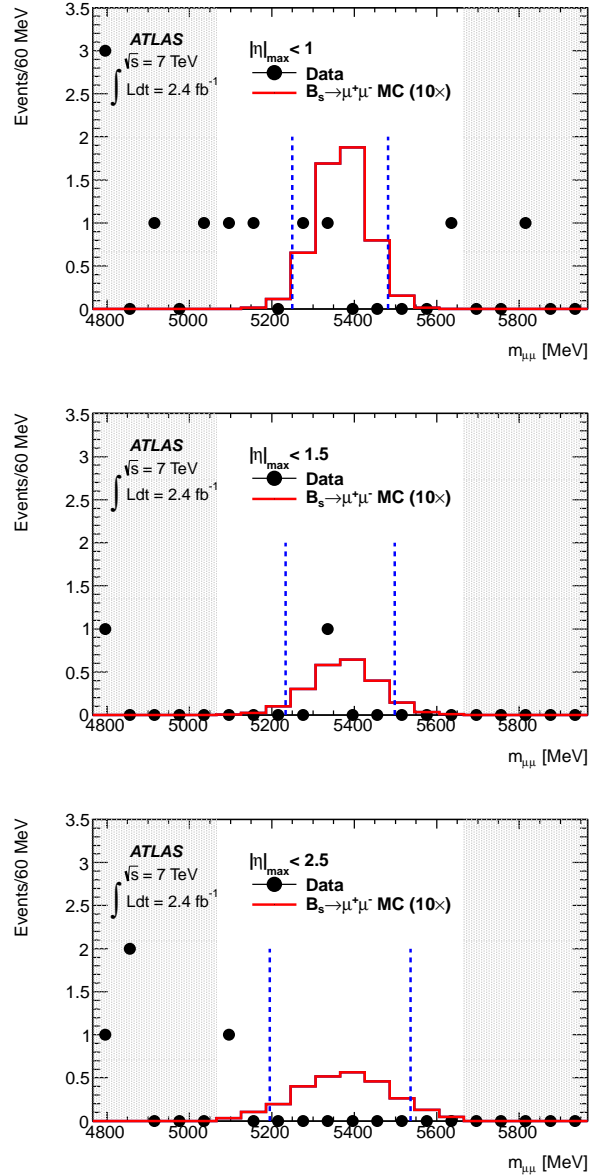}
\caption{\label{fig:UnblindedMass} Invariant mass distribution of candidates in data. For each mass-resolution category (top to bottom) each plot shows the invariant mass distribution for the selected candidates
in data (dots), the signal (continuous line) as predicted by MC assuming ${\rm BR}(\Bs \to \mumu)=3.5\cdot 10^{-8}$, and two dashed vertical lines corresponding to the optimized $\Delta m$ cut. The grey areas
correspond to the sidebands used in the analysis.}
\end{center}
\end{figure}

The values of the SES 
are given in Table~\ref{table:SES-counts} which also shows the values of the parameters $\thetau_i$, the background counts in the sidebands\footnote
{For comparison, the number of odd-numbered events observed in the sidebands, which is expected to be biased due to the use of the same sample in selection optimization and BDT training, was found to be equal to one event in each of the three mass-resolution categories.},
the resonant background, and finally the observed number of events in the search region, as found after unblinding.  Figure \ref{fig:UnblindedMass} shows the invariant mass distribution
of the selected candidates in data, for the three mass categories, together with the signal projections as obtained from MC assuming ${\rm BR}(\Bs \to \mumu)=3.5\cdot 10^{-8}$ (i.e. approximately $10$ times the
SM expectation).
 
\section{Branching fraction limits}\label{sec:limits}

The upper limit on the \Bs$ \to\! \mu^+\! \mu^-$ branching fraction was obtained by means of an implementation 
\cite{Junk:1999kv} of the CL$_\mathrm s$ method \cite{Read:2002hq}.  The extraction was based on the likelihood:
\begin{eqnarray*}
\hbox{\hspace{-8mm}}\mathcal{L}\hbox{\hspace{-2mm}}&=& \hbox{\hspace{-2mm}} {\rm Gauss}(\epsilon_{\rm obs}|\epsilon,\sigma_\epsilon)\times {\rm Gauss} (\thetau_{\rm obs}|\thetau,\sigma_{\thetau}) \times \\
&  &\hbox{\hspace{-2mm}} \prod_{i=1}^{N_{\rm bin}} {\rm Poisson}(N^{\rm obs}_i|\epsilon\,\epsilon_i {\rm BR} + N^\mathrm {bkg}_i +N^{B\to hh}_i) \times \\
& &\hbox{\hspace{-2mm}} {\rm Poisson}(N^\mathrm {bkg}_{{\rm obs},i}|\thetau\,\thetau_i N^\mathrm {bkg}_i) \times \\ 
&  &{\rm Gauss}(\epsilon_{{\rm obs},i}|\epsilon_i,\sigma_{\epsilon_i})\, .
\label{eqn:Likelihood}
\end{eqnarray*}
For each mass-resolution category, the likelihood contains Poisson distributions for the event counts in the search and sideband regions and a Gaussian distribution for the relative efficiency $\epsilon_i$. Two additional Gaussians describe the coherent systematic uncertainties in $\thetau$ and in the SES. The mean of the Poisson distribution in the search region is equal to the sum of the \Bs\ branching fraction (scaled by the normalization and relative efficiency parameters), the continuum background and the resonant background. The mean of the Poisson distribution in the sidebands is equal to the background scaled by $\thetau$. The parameters $\sigma_\epsilon$ ($\sigma_{\epsilon_i}$), $\sigma_{\thetau}$ ($\sigma_{\thetau_i}$) account for the correlated (uncorrelated) uncertainties in the SES and the background scaling factor.
In this analysis the uncertainties on $\thetau_i$ are negligible, with $\thetau = 1.00 \pm 0.04$. All input parameters are summarized in Table~\ref{table:SES-counts}.

\begin{figure}[b!]
\begin{center}
\includegraphics[width=\linewidth]{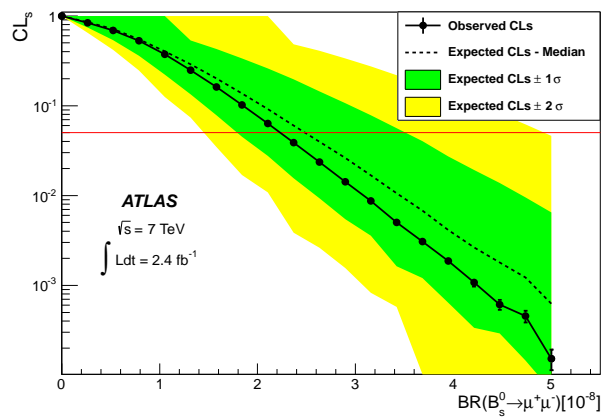}
\caption{\label{fig:FinalCLs} Observed CL$_\mathrm s$ (circles) as a function of BR($\Bs \to\! \mu^+\! \mu^-$). The 95\% CL limit is indicated by the horizontal (red) line. The dark (green) and light (yellow) bands correspond to $\pm 1\sigma$ and $\pm 2\sigma$ fluctuations on the expectation (dashed line), based on the number of observed events in the signal and sideband regions.}
\end{center}
\end{figure}

The expected limits were obtained by setting the counts in the search region equal to the interpolated background plus the small resonant background, 
before the unblinding of the signal region. A median expected limit of
 $2.3^{+1.0}_{-0.5} \times 10^{-8}$  at 95\% CL was obtained, where the range encloses 68\% of the background-only pseudo-experiments.

For comparison the mass-resolution categories were merged and the selection optimization was performed on the merged sample. In this case eight events were found in the sidebands, resulting in a branching fraction limit of $2.9^{+1.3}_{-0.8} \times 10^{-8}$ at 95\% CL. This test confirms the expectation of a more sensitive analysis when separate mass-resolution categories are used.

The background counts found in odd-numbered events were used to assess the magnitude of the bias that would be caused by using the same sample for selection optimization and the estimation of $N^{\rm bkg}$. The expected limit obtained using the same sample for optimization and signal extraction is $1.7 \times 10^{-8}$, about 30\% smaller than the limit presented in this Letter, for which independent samples were used for optimization and for signal extraction. The observed bias is consistent with simulation-based assessments of this effect.

Figure~\ref{fig:FinalCLs} shows the behaviour of the observed  
CL$_\mathrm s$ for different tested values of the \Bsmumu\ branching fraction, computed with 300\,000 toy MC 
simulations per point. The observed limit is $<2.2$ (1.9) $\times 10^{-8}$  at 95\% (90\%) CL. The {\em p\/}-values for the background-only hypothesis and for background plus SM prediction \cite{Buras:2010zm, Buras:2010wr} are 44\% and 35\%, respectively. 

Despite the difference between the total numbers of observed and interpolated background events (equal to 3 and 6.5, respectively), the interplay of the event counts observed in the three mass resolution categories produced an observed CL$_\mathrm s$ limit close to the expected value.  

\section{Conclusions}
A limit on the branching fraction BR($\Bs \to\! \mu^+\! \mu^-$) is set using 
2.4~fb$^{-1}$ of integrated luminosity collected in 2011 by the ATLAS detector. 
The process $B^\pm\to J/\psi K^\pm$, with $J/\psi \to \mumu$, is used as a reference channel 
for the normalization of integrated luminosity, acceptance and efficiency. The final selection is based on a multivariate analysis performed on three categories of events determined according to their mass resolution, yielding a limit of  BR($\Bs \to\! \mu^+\! \mu^-) < 2.2 \, (1.9)\times 10^{-8}$  at 95\% (90\%) CL.


\section{Acknowledgements}

We thank CERN for the very successful operation of the LHC, as well as the
support staff from our institutions without whom ATLAS could not be
operated efficiently.

We acknowledge the support of ANPCyT, Argentina; YerPhI, Armenia; ARC,
Australia; BMWF, Austria; ANAS, Azerbaijan; SSTC, Belarus; CNPq and FAPESP,
Brazil; NSERC, NRC and CFI, Canada; CERN; CONICYT, Chile; CAS, MOST and NSFC,
China; COLCIENCIAS, Colombia; MSMT CR, MPO CR and VSC CR, Czech Republic;
DNRF, DNSRC and Lundbeck Foundation, Denmark; EPLANET and ERC, European Union;
IN2P3-CNRS, CEA-DSM/IRFU, France; GNAS, Georgia; BMBF, DFG, HGF, MPG and AvH
Foundation, Germany; GSRT, Greece; ISF, MINERVA, GIF, DIP and Benoziyo Center,
Israel; INFN, Italy; MEXT and JSPS, Japan; CNRST, Morocco; FOM and NWO,
Netherlands; RCN, Norway; MNiSW, Poland; GRICES and FCT, Portugal; MERYS
(MECTS), Romania; MES of Russia and ROSATOM, Russian Federation; JINR; MSTD,
Serbia; MSSR, Slovakia; ARRS and MVZT, Slovenia; DST/NRF, South Africa;
MICINN, Spain; SRC and Wallenberg Foundation, Sweden; SER, SNSF and Cantons of
Bern and Geneva, Switzerland; NSC, Taiwan; TAEK, Turkey; STFC, the Royal
Society and Leverhulme Trust, United Kingdom; DOE and NSF, United States of
America.

The crucial computing support from all WLCG partners is acknowledged
gratefully, in particular from CERN and the ATLAS Tier-1 facilities at
TRIUMF (Canada), NDGF (Denmark, Norway, Sweden), CC-IN2P3 (France),
KIT/GridKA (Germany), INFN-CNAF (Italy), NL-T1 (Netherlands), PIC (Spain),
ASGC (Taiwan), RAL (UK) and BNL (USA) and in the Tier-2 facilities
worldwide.

\bibliography{BMuMu_PLB}
\bibliographystyle{elsarticle-num}
\nocite{*}

\clearpage
\onecolumn
\begin{flushleft}
{\Large The ATLAS Collaboration}

\bigskip

G.~Aad$^{\rm 48}$,
B.~Abbott$^{\rm 111}$,
J.~Abdallah$^{\rm 11}$,
S.~Abdel~Khalek$^{\rm 115}$,
A.A.~Abdelalim$^{\rm 49}$,
O.~Abdinov$^{\rm 10}$,
B.~Abi$^{\rm 112}$,
M.~Abolins$^{\rm 88}$,
O.S.~AbouZeid$^{\rm 158}$,
H.~Abramowicz$^{\rm 153}$
% ...
\bigskip

% ...
$^{10}$ Institute of Physics, Azerbaijan Academy of Sciences, Baku, Azerbaijan\\
$^{11}$ Institut de F\'isica d'Altes Energies and Departament de F\'isica de la Universitat Aut\`onoma  de Barcelona and ICREA, Barcelona, Spain\\
% ...
$^{48}$ Fakult\"{a}t f\"{u}r Mathematik und Physik, Albert-Ludwigs-Universit\"{a}t, Freiburg i.Br., Germany\\
$^{49}$ Section de Physique, Universit\'e de Gen\`eve, Geneva, Switzerland\\
% ...
$^{88}$ Department of Physics and Astronomy, Michigan State University, East Lansing MI, United States of America\\
% ...
$^{111}$ Homer L. Dodge Department of Physics and Astronomy, University of Oklahoma, Norman OK, United States of America\\
$^{112}$ Department of Physics, Oklahoma State University, Stillwater OK, United States of America\\
% ...
$^{115}$ LAL, Universit\'e Paris-Sud and CNRS/IN2P3, Orsay, France\\
% ...
$^{153}$ Raymond and Beverly Sackler School of Physics and Astronomy, Tel Aviv University, Tel Aviv, Israel\\
% ...
$^{158}$ Department of Physics, University of Toronto, Toronto ON, Canada\\
% ...
$^{*}$ Deceased\end{flushleft}


\end{document}